\documentstyle[editedvolume,numreferences,epsf]{crckapb}

\begin{opening}
\title{Admittance and nonlinear transport 
in quantum wires, point contacts, and resonant tunneling barriers}
\author{M.\ B\"uttiker and T. Christen}
\institute{Department of Theoretical Physics,\\
University of Geneva, \\ 24, Quai E. Ansermet,  CH-1211 Geneva 4, Switzerland}   
\end{opening}
\runningtitle{Admittance and nonlinear transport}
\begin{document}

\section{Introduction}
\label{introduction}

The admittance determines the current response of an electrical conductor to oscillating voltages
applied to its contacts. The admittance and the nonlinear dc-transport have in common that 
they require both an understanding of the charge distribution which is 
established as the conductor is driven away from its equilibrium state. The dynamic
and the nonlinear transport can thus be viewed as probes of the charge response of the conductor.
In contrast, linear dc-transport is determined by the equilibrium charge distribution 
alone. Thus the dynamic conductance and the nonlinear transport provide information
which cannot be gained from a linear dc-measurement. 

It is the purpose of this article to provide a discussion of ac-transport and nonlinear 
transport in small mesoscopic conductors. 
Our discussion leans heavily on the scattering approach to electrical dc-conductance and 
can be viewed as an extension of this approach to treat transport beyond the 
stationary ohmic regime. Our emphasis is on the low-frequency admittance for which 
expressions can be derived which are very general and can be applied to 
a large class of problems. Similarly our discussion emphasizes the departure from 
linear ohmic transport in the case when nonlinearity first becomes apparent. 
The principles which govern the low-frequency behavior and the weakly nonlinear 
behavior apply of course as well for high-frequency transport and for strongly
nonlinear transport. We underline this fact by treating the simple example of 
a resonant tunneling barrier for its entire frequency range 
and by calculating the nonlinear I-V characteristic 
for a large range of applied voltages. 

The general guiding principles of our discussion are charge and current conservation
and the need to obtain gauge invariant expressions. The conservation of charge becomes 
a fundamental requirement under the following condition \cite{BUTT93d}: 
Suppose that it is possible to 
find a volume $\Omega$ which encloses the mesoscopic part of the conductor including a portion
of its contacts such that all electric field lines remain within this volume. 
Clearly, if the mesoscopic conductor is formed with the help of gates the volume 
$\Omega$ must be large enough to enclose also a portion of the gates. 
The electric field is then localized within $\Omega$. Such localized electric (and magnetic)
field distributions are always implicitly assumed when a circuit is represented 
simply in terms of $R$, $C$ and $L$ elements. If the field is localized in $\Omega$
the electric flux through the surface of that volume is zero and according to the law
of Gauss the total charge inside this volume is zero. Consequently, application of a 
time-dependent or stationary voltage to the conductor (or a nearby gate)
leads only to a redistribution of charge within $\Omega$ but leaves the overall charge invariant.
Therefore, a basic ingredient of a reasonable discussion of transport of an electrical
conductor is a theory of the charge distribution with the overall constraint that the 
total charge is conserved. 
If the total charge is conserved then the currents measured at the contacts of the sample 
are also conserved. As we will show, this leads to sum rules for the 
admittance coefficients of a mesoscopic conductor and leads to sum rules
for the coefficients which govern the nonlinear dc-transport. 
The conservation of the total charge is also connected with the fact that an electrical conductor
does not change its properties if we shift all potentials by an equal amount.
The principle of gauge invariance states that the results can depend only on 
voltage differences.  This leads to a second set of sum rules for the admittance coefficients 
and for the nonlinear transport coefficients \cite{CHRI96c}. 

The three guiding principles which we have taken for the development of a theory 
are thus very closely related and hinge on the localization of electric and magnetic fields.
The existence of a Gauss volume is not, however, a trivial requirement. A different point of view
is taken in a recent article by Jackson \cite{JACK}. In his circuit the potential distribution
depends on the location of the battery vis-\`a-vis the resistor. Such a point of view 
would make it necessary to develop a theory for the entire circuit 
including its components used to drive it out of equilibrium. We believe that the notion
of localized fields is a more fruitful one and is closer to the experimental reality
as it is encountered in mesoscopic physics. 
In what follows, it is not only assumed that the electric fields are localized, 
but that the localization volume $\Omega$ is in fact of the same dimension as a 
phase breaking length. This permits us to formulate a theory for the mesoscopic
structure alone and to treat the completion of the circuit as a macroscopic problem. 

A charge and current conserving approach to mesoscopic ac-conductance was 
developed by one of the authors in collaboration with 
Pr\^etre and Thomas \cite{BUTT93b}. Similar work, without an effort to achieve 
a self-consistent description, was also presented by Pastawski \cite{PAST92a} and 
Fu and Dudley \cite{FUYD93a}.
In Ref. \cite{BUTT93b} charge and current conservation was achieved by attributing a 
{\it single} 
self-consistent potential to a conductor. A discussion which treats the potential 
{\it landscape}
in a Hartree-like approach was given in Ref. \cite{BUTT94a} for metallic screening and 
in Ref. \cite{BUTT93d} for long range screening appropriate for semiconductors.
These two discussions can be extended to include an effective 
potential as it occurs in density functional theory\cite{BUTT95b}. Both the limit 
of a single potential and the case of a continuous potential are important but for 
many applications the first is to crude and the later is to complex. In this work 
we will treat an approach which is in between these two: In this generalized discrete 
potential model the conductor is divided into as many regions as is necessary 
to capture the main features of the charge and potential distribution. 
This permits the description of charge distributions which are mainly dipolar
or of a more complex higher order multi-polar form. 
The authors 
applied such an approach to 
the low-frequency behavior of quantized Hall samples \cite{CHRI96a}, 
quantum point contacts \cite{CHRI96b} and to nonlinear transport \cite{CHRI96c}.

We will only briefly review the scattering approach to electrical conductance, mainly 
to introduce the notation used in this article. For a more extended discussion of some
of the basic elements of the scattering approach we refer the reader to the introductory 
chapter of this book or to one of the reviews \cite{IMRY86a,BEEN91a,BUOT,DATT93a}. 
The advantage of the scattering approach, as formulated by Landauer \cite{LAND70a},
Imry\cite{IMRY86a} and one of the authors \cite{BUTT86a,BUTT88b} 
lies in the conceptually simple
prescription of how to model an {\em open} system, 
i.e. how to couple the sample to external contacts
which act as reservoirs of charge carriers and provide the source of irreversibility.
The self-consistent potential in the context of transport was already of interest 
to Landauer \cite{LAND57a} and is in fact the main
issue which distinguishes his early work from preceding works on
transmission through tunnel contacts
\cite{FREN30a}. 
A more recent discussion of the potential distribution 
in ballistic mesoscopic conductors is provided by Levinson \cite{LEVI89a}
and technically is closely related to our approach \cite{BUTT93d}.
For the discrete potential model discussed here, the potential distribution
can be discussed in a purely algebraic formulation which we will present in some detail.
A large portion of this article is devoted to an explicite illustration of our 
approach. We discuss the low frequency admittance of wires with barriers and 
the low frequency admittance of quantum point contacts. In particular, we generalize 
published work on the low frequency admittance of the quantum point contact 
to investigate the effect of the gates used to form the quantum point contact. 
The last section of this work presents a discussion of ac-transport and nonlinear 
transport over a large frequency and voltage regime under the assumption
that the only important displacement current is that to a nearby gate.

The discussion of potentials also sheds light on the limits of validity of the 
scattering approach as it is used to describe dc-transport: 
The scattering approach is often classified as a noninteracting theory:
That is incorrect. Since scattering states can be calculated in an effective potential
the range of validity of the scattering approach is at least the same 
as density functional 
theory. That makes it possible to include 
exchange and correlation effects and  
makes the scattering approach a mean-field theory. 
As is known, 
for instance from the BCS-theory of superconductivity, mean-field theories can be
very powerful tools which render any remaining deviations very hard to detect. 

The low-frequency behavior discussed here should experimentally be 
accessible with much of the same techniques as those that are used
for dc-measurements. We mention here the work of
Chen et al. \cite{CHEN94a} and Sommerfeld et al. \cite{SOMM96a}
on the magnetic field 
symmetry of capacitances. We mention further a recent experiment by 
Field et al. \cite{FIEL96a} who used 
a quantum dot capacitively coupled to a two-dimensional electron gas 
to measure the density of states at the metal-insulator transition.
In contrast, experiments at high-frequencies require 
special efforts to couple the signal to the mesoscopic conductor 
and to measure the response. As examples,
we cite here the work by Pieper and Price on the admittance of an Aharonov-Bohm 
ring \cite{PIEP94a},
investigations of photo-assisted tunneling by Kouwenhoven et al. \cite{KOUW94a}, 
noise measurements at large frequencies \cite{REZN95a}, 
and transport in superlattices \cite{HOFB}. 
We emphasize that already the low-frequency 
linear admittance presents a very interesting characterization of the 
sample. Indeed, we hope, that this article demonstrates that there is considerable
room for additional theoretical and experimental work in this area.\\   
Nonlinear dc-effects of interest are asymmetric current-voltage 
characteristics and rectification
\cite{TABO94a}, the evolution
of half-integer conductance plateaus \cite{GLAZ89a,PATE90a}, the breakdown of
conductance quantization \cite{KOUW89a}, and negative differential
conductance and hysteresis \cite{KLUK89a}.
Like in frequency-dependent transport most of the theoretical work shows no
concern for self-consistent effects. An exception is a numerical 
treatment of a tunneling barrier by Kluksdahl et al. \cite{KLUK89a}.\\
The emphasis on the role of the long-range Coulomb interaction 
in this work should be contrasted with the recent discussions 
of mesoscopic transport in the framework of the Luttinger approach. 
The Luttinger approach treats the short-range interactions only.
However, since a one-dimensional wire cannot screen
charges completely, a more realistic treatment should include 
the Coulomb effects. Without the long-range Coulomb interaction
the results based on Luttinger models only are not charge and current conserving and
are not gauge invariant. In this context we note that the theory of the Coulomb
blockade has been very successful in describing the important experimental aspects.
The Coulomb blockade is a consequence of long-range and not of  
short-range interactions. In any case the Hartree discussion given here
is useful as a comparison with other theories which include interactions.
To distinguish non-Fermi liquid behavior from those of a Fermi liquid it will 
be essential to compare experimental data with a reasonable, i. e. self-consistent, 
Fermi liquid theory.

%
\section{Theory}
\label{Theory}
%
In this section we recall briefly the theory of ac-transport and weakly nonlinear 
transport which has
been developed in Refs. \cite{BUTT93d,CHRI96c,BUTT93b,BUTT94a,CHRI96b}.
A review of these works is provided by Ref. \cite{BUTT96a}.\\

%
\subsection{Mesoscopic Conductors}
\label{MC}
\begin{figure}[t]
\hspace{0mm}
 \epsfxsize = 50 mm
 \epsffile{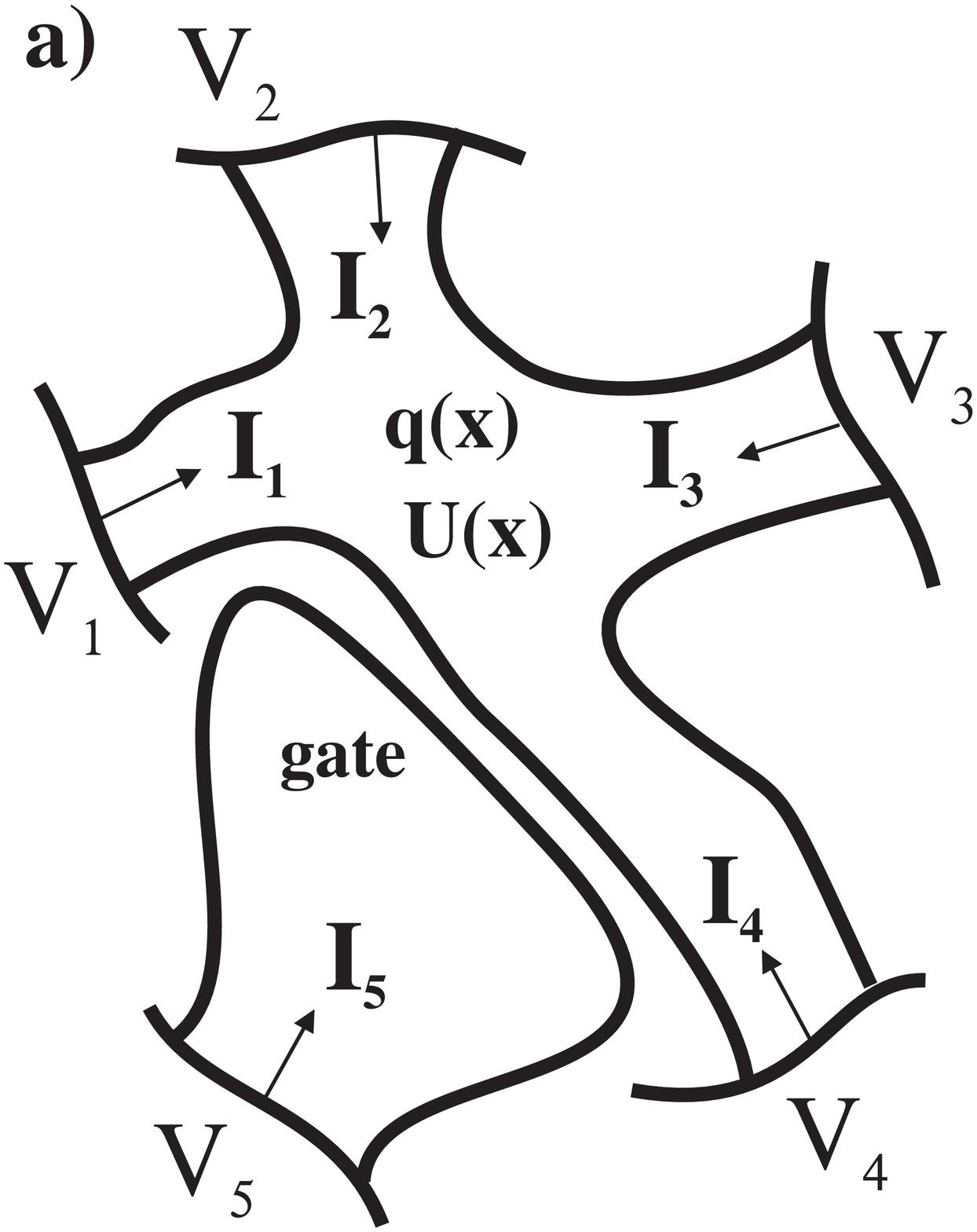}
\hspace{15mm}
 \epsfxsize = 50 mm
 \epsffile{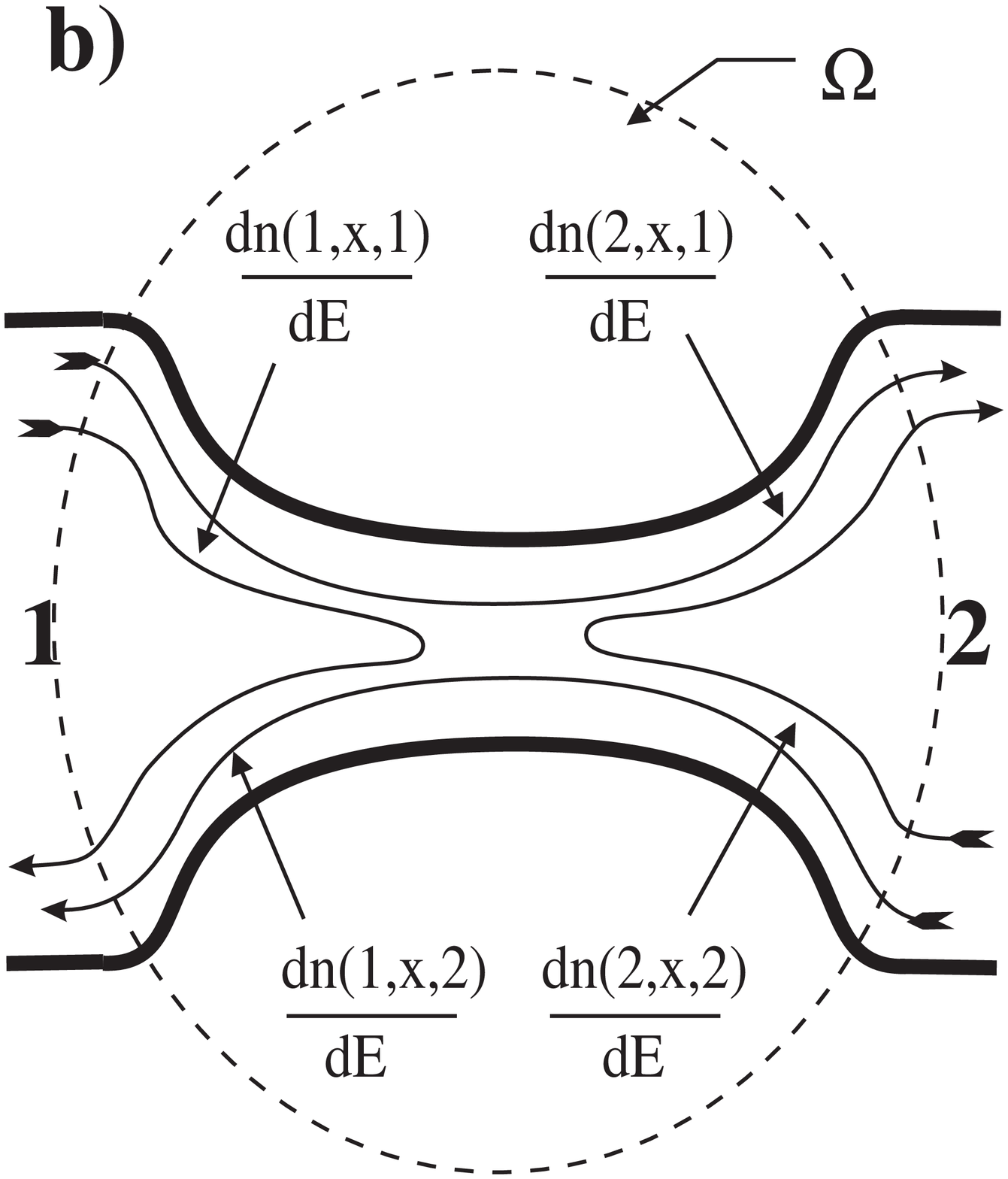}
\caption{a) Multi-terminal sample including a gate; the voltages $V_{\alpha}$
induce charges ($q(x)$) which influence the potential ($U(x)$),
and  drive currents ($I_{\beta}$). b)  
Decomposition of the local density of states in a two-terminal sample into local
partial densities of states.}
\label{fig1}
\end{figure}

%
Consider a conducting region
which is connected via leads and contacts $\alpha = 1,...,N$ 
to $N$ reservoirs of charge carriers as shown in Fig. \ref{fig1}a.
In order to include into the formalism the presence of nearby
gates (capacitors), some parts of the sample are allowed to be macroscopically large
and disconnected from other parts. A magnetic
field $B$ may be present. It is assumed that the leads of the conducting part
are so small that carrier motion occurs via one-dimensional subbands (channels)
$m =1, ... , M_{\alpha}$ for each contact. Moreover,
the distance between these contacts is assumed to be short compared to
the inelastic scattering length and the phase-breaking length, such that transmission
of charge carriers from one contact to another one can be considered
to be elastic and phase coherent.  A reservoir is at
thermodynamic equilibrium and can thus be characterized by its 
temperature T$_{\alpha}$ and by
the electrochemical potential $\mu _{\alpha}$ of the carriers. While we assume that all reservoirs
are at the same temperature T, we consider differences of the
electrochemical potentials which is usually achieved by applying voltages $V_{\alpha}$ to the
reservoirs. We fix the voltage scales such that  
$V_{\alpha} \equiv 0$ corresponds to the
equilibrium reference state 
where all electrochemical potentials of the reservoirs are equal to each other,
$\mu _{\alpha} \equiv \mu_{0}$. The problem to be solved is now to find the current
$I_{\alpha}(t)$ entering the sample at contact $\alpha $ in response 
to a (generally time
dependent) voltage $V_{\beta}(t)$ at contact $\beta $. The
general time-dependent nonlinear transport problem is highly nontrivial
and we consider mainly {\it weakly} nonlinear dc-transport and {\it linear low}-frequency 
transport.\\ 

%
%
\subsection{The scattering approach}
\label{SA}
The sample is described by a unitary
scattering-matrix ${\bf S}$ \cite{IMRY86a,LAND70a,BUTT86a,BUTT85a}. 
Unitarity together with micro-reversibility 
implies that under a reversal of the magnetic field the scattering matrix has
the symmetry ${\bf S}^{T}(B) = {\bf S} (-B)$ \cite{BUTT86a,BUTT88b}. 
Furthermore, the scattering
matrix ${\bf S}$ for the conductor can be arranged such that
it is composed of sub-matrices ${\bf s}_{\alpha \beta }(E,[U(x)]) $ with
elements $s_{\alpha \beta n m } (E,[U(x)])$ which relate the out-going current amplitude
in channel $n$ at contact $\alpha $ to the incident current amplitude in channel
$m$ at contact $\beta $. The scattering matrix is a function of the
energy $E$ of the carrier and is a functional of the electric potential $U(x)$
in the conductor \cite{BUTT93d}. For
the dc-transport properties, all the physical information which is needed, 
is contained in the scattering matrix.\\
The electric potential $U(x,\{V_{\gamma}\})$ is a function
of the voltages $V_{\gamma} = (\mu_{\gamma} - \mu_{0})/e$ applied at the 
contacts and at the nearby gates.
Thus the scattering matrix depends not only on the energy of the scattered carriers
but also on the voltages, ${\bf s}_{\alpha \beta}(E, \{  V_{\gamma}\})$.
With the help of the scattering matrix, we can find the current $dI_{\alpha}(E)$
in contact $\alpha$ due to carriers incident at contact $\beta $ in the energy 
interval $(E, E+dE).$ It is convenient to introduce a {\it spectral} 
conductance \cite{CHRI96c} 
\begin{equation}
g_{\alpha \beta}(E) = \frac{e^{2}}{h}\:
Tr[ {\bf 1}_{\alpha } (E,\{ V_{\gamma}\})\delta _{\alpha \beta}
-{\bf s}^{\dagger}_{\alpha \beta }(E,\{ V_{\gamma}\})
{\bf s}_{\alpha \beta }(E,\{ V_{\gamma}\})]  
\label{spectralg}
\end{equation}
such that the current at energy $E$ becomes 
$dI_{\alpha}(E) = g_{\alpha \beta} (E) (dE/e) .$  
The unity matrix ${\bf 1}_{\alpha }$ lives
in the space of the quantum channels in lead $\alpha $ with thresholds below
the electrochemical potential. Note that
this matrix is also a (discontinuous) function of energy and of the
potential. It changes its dimension by one whenever the band bottom
of a new subband passes the electrochemical potential. 
The current through contact $\alpha $ is the sum of all spectral currents
weighted by the Fermi functions $f(z)=(1+\exp (z/k_{B}{\rm T}))^{-1}$ of the
reservoirs \cite{BUTT92b} at temperature T
\begin{equation}
I_{\alpha}= 
\sum _{\beta = 1} ^ {N}
\int (dE/e)\: f(E- \mu_{0} - eV_{\beta}) \:
g_{\alpha \beta}(E,\{ V_{\gamma}\}) \;\; .
\label{current}
\end{equation}
Linear transport is determined by an expansion of Eq. (\ref{current})
away from the equilibrium reference state to linear order in $V_{\gamma}$.
The linear conductance is found to be
\begin{equation} 
G_{\alpha \beta}^{(0)}= \int dE\:(-\partial _{E}f_{0})\; 
g_{\alpha \beta}(E) \;\; ,
\label{dccond}
\end{equation}
where the $G_{\alpha \beta} $
are evaluated at $V_{1}=...=V_{N}= 0$.
Consequently, the linear conductances depend on the 
equilibrium electrostatic potential $U^{(eq)}(x) \equiv U(x,\{ 0 \})$  only.
In contrast, both the ac-transport and the nonlinear transport
depend explicitly on the nonequilibrium potential.
A discussion of the nonequilibrium potential requires 
knowledge of the local charge distribution in the conductor.
Within linear response, the charge response is related to  
the DOS of the conductor at the Fermi energy. 
The local DOS 
can be obtained from the scattering matrix by a functional derivative 
with respect to the local potential \cite{BUTT93d}
\begin{eqnarray}
\frac{dn(x)}{dE} =
- \frac{1}{4 \pi i} \sum_{\alpha \beta}
Tr\left[{\bf s}^{\dagger}_{\alpha\beta}
  \frac{\delta {\bf s}_{\alpha\beta}}{e\delta U(x)}
- \frac{\delta {\bf s}^{\dagger}_{\alpha\beta}}{e\delta U(x)}
{\bf s}_{\alpha\beta} \right] 
\equiv \sum_{\alpha \beta}\frac{dn (\alpha , x, \beta )}{dE} \;\; .
\label{LDOS}
\end{eqnarray}
As indicated in Eq. (\ref{LDOS}), the local DOS
can be understood as a sum of local {\it partial densities of states} (PDOS) 
$dn(\alpha ,x ,\beta )/dE$ \cite{GASP96a}.
The sum is over all injecting contacts $\beta $ and all emitting contacts $\alpha $. 
The meaning
of a local PDOS $dn(\alpha ,x ,\beta )/dE$ is then obvious: 
it is the local density of states associated with
carriers which are scattered from contact $\beta $ to contact $\alpha $. 
The local PDOS are illustrated in Fig. \ref{fig1}b. 
The global DOS, $dN/dE$, follows by a spatial integration of Eq. (\ref{LDOS}) 
over the sample region $\Omega$
which encloses the mesoscopic structure with a boundary deep inside the reservoirs.
Integration of the local PDOS over the whole sample 
leads to the global PDOS, $dN_{\alpha \beta}/dE$.
Clearly, it holds $dN/dE = \sum _{\alpha \beta} dN_{\alpha \beta}/dE $.
Note that Eq. (\ref{LDOS}) counts only scattering states. 
Pure bound states, e.g. trapped or pinned 
at an impurity, are not included.\\
The local PDOS $dn(x ,\beta )/dE$ which contains information
only on the contact from which the carriers enter the conductor
(irrespectively of the contact through which the carriers exit)
is called the {\em injectivity} of contact $\beta $
and is given by $dn(x ,\beta )/dE = \sum_{\alpha} dn(\alpha ,x ,\beta )/dE$. 
The local PDOS $dn(\alpha ,x)/dE$ which contains 
information only on the contact through which carriers are leaving the sample
but contains no information on the contact through which 
carriers entered the conductor is called the {\em emissivity}
of contact $\alpha$ and is given by
$dn(\alpha ,x)/dE = \sum_{\beta} dn(\alpha ,x ,\beta )/dE$.
Due to reciprocity, the injectivity at a magnetic field $B$
is equal to the emissivity at the reversed magnetic field,
$dn_{B}(x ,\alpha)/dE = dn_{-B}(\alpha ,x)/dE $.\\
Throughout this work we use the discrete potential approximation
in which the conductor is divided into $M$ regions $\Omega_{k}$,
$k = 1,  ..., M$.
The charge and the potential in region $k$ are 
denoted by $q_{k}$ and $U_{k}$. Since we are 
are interested in the charge variation in response 
to a variation in voltage, we introduce the DOS of
$\Omega_{k}$. For later convenience, we express these DOS in units 
of a capacitance by multiplication with $e^{2}$. 
Thus, $D_{k} = e^{2} \int _{\Omega_{k}} dx\: (dn(x)/dE)$ 
is the DOS of region $k$. The PDOS of region $k$ are 
$D_{\alpha k \beta} = e^{2} \int _{\Omega_{k}} dx\: (dn(\alpha ,x ,\beta )/dE)$.
Obviously, the injectivities and the emissivities are 
$D_{k \beta} = e^{2} \int  _{\Omega_{k}} dx\: (dn(x ,\beta )/dE)$ and
$D_{\alpha k } = e^{2} \int _{\Omega_{k}} dx\: (dn(\alpha , x )/dE)$, respectively.


\subsection{Linear low-frequency transport}
\label{LACT}
We are interested in the admittance
$G_{\alpha\beta} (\omega) $ which determines the Fourier amplitudes
of the current, $\delta I_{\alpha}(\omega) $,
at a contact $\alpha$ in response to an oscillating voltage 
$\delta V_{\beta}(\omega) \exp(-i\omega t)$ at contact $\beta$
\begin{equation}
\delta I_{\alpha} = \sum _{\beta}
G_{\alpha \beta} (\omega) \; \delta V_{\beta}\;\;.
\label{admittance}   
\end{equation} 
To investigate the low-frequency limit, we expand the admittance in powers of frequency
\begin{equation}
G_{\alpha \beta} (\omega)= G_{\alpha \beta} ^{(0)}-i\omega E_{\alpha \beta}
+ \omega^{2} K_{\alpha \beta}  + {\cal O}(\omega ^{3}) \;\; .
\label{expansion} 
\end{equation}
The dc-conductance $G_{\alpha \beta} ^{(0)}$ has already been derived in Eq. (\ref{dccond}).
The first order term is determined by the \em emittance \rm matrix $E_{\alpha \beta}
\equiv i(dG_{\alpha \beta}/d\omega)_{\omega = 0}$.
The emittance governs the displacement currents of the mesoscopic structure. 
The second order term,  $K_{\alpha \beta}$, contains information on the 
charge relaxation of the system. As an example, consider for a moment
a macroscopic capacitor $C$ in series
with a resistor $R$. For this purely capacitive structure with vanishing
dc-conductance, $G=0$, the emittance is the
capacitance, $E = C$, and $K=R^{(q)}C^{2}$ contains the
$RC$ time with the charge relaxation resistance $R^{(q)} \equiv R$.
More generally, for any structure consisting of $N$ metallic conductors,
each connected to a single reservoir,  
the emittance matrix is just the capacitance matrix.
These simple results must be modified for {\em mesoscopic} 
conductors and conductors which
connect {\em different} reservoirs.
Firstly, it is not the geometrical capacitance but rather the 
DOS dependent {\em electro-chemical
capacitance} which relates charges
on mesoscopic conductors to voltage variations
in the reservoirs \cite{BUTT93a}. Secondly, conductors which connect
different reservoirs allow a transmission of charge which leads to
inductance-like contributions to the emittance \cite{BUTT93d,BUTT93b,CHRI96b}.
Thirdly, the charge relaxation resistance cannot, in general,
be calculated like a dc-resistance  \cite{BUTT93a,BUTT96b}.\\
To illustrate our approach, we derive now a general expression for the emittance. We first
notice that for the purely capacitive case the current (displacement current) at contact $\alpha $
is the time derivative of the total charge $\delta Q_{\alpha}$ on the capacitor plates
connected to this contact. Hence, $\delta I_{\alpha}= -i\omega \delta Q_{\alpha}$
and the emittance is given by $E_{\alpha \beta}= \partial Q_{\alpha}/\partial V_{\beta}$.
If transmission between different reservoirs is allowed, the charge $\delta Q_{\alpha}$
scattered through a contact can no longer be associated with a unique spatial region.
The charge at a given point is rather injected at different contacts and 
ejected at different contacts. However, 
the charge emitted at a contact can still be calculated within
the scattering approach. We decompose it in a
bare and a screening part 
$ \delta Q_{\alpha} = \delta Q_{\alpha}^{(b)} + \delta Q_{\alpha}^{(s)} $.
The bare part of the charge corresponds to the charge which is scattered through
the contact $\alpha $ for fixed electric potential and is thus given by
\begin{equation}
\delta Q _{\alpha} ^{(b)} =  \sum _{\beta} D_{\alpha \beta} \delta V_{\beta} \;\; ,
\label{qb2}
\end{equation}
where $D_{\alpha \beta} = \sum_{k} D_{\alpha k \beta}$
is the global PDOS. The screening part, on the other hand, 
is associated with the charge which is scattered through
contact $\alpha $ for a variation in the electric potential only. Since a shift of the band
bottom contributes with a negative sign and since the potential is in general spatially
varying, $ \delta Q_{\alpha}^{(s)} $ is connected to the local PDOS by
$\delta Q _{\alpha}^{(s)} = - \sum_{k} D_{\alpha k} \delta {U}_{k},$
where $D_{\alpha k}$ is the emissivity of region $k$ into contact $\alpha$.
If we introduce the vector of emissivities ${\bf D}^{e}_{\alpha} = (D_{\alpha 1}, .., D_{\alpha M})$
and write the potential variation as a vector, $\delta {\bf U} = (\delta {U}_{1}, ...., \delta {U}_{M})$,
the charge emitted into contact $\alpha$ is  
\begin{equation}
\delta Q _{\alpha}^{(s)} = -  {\bf D}^{e}_{\alpha} \delta {\bf U}\;\; .
\label{qs2}
\end{equation}
In linear response the potential variation $\delta {\bf U}$ 
is linearly related to the potential variations $\delta V_{\beta}$ at the 
contacts of the sample. We write 
\begin{eqnarray}
\delta {\bf U} = \sum_{\alpha} {\bf u}_{\alpha} \: \delta V_{\alpha}\;\;
\label{charfu}
\end{eqnarray}
where the response coefficients ${\bf u}_{\alpha}= (u_{1\alpha} , ..., u_{M\alpha})$ are 
called the {\it characteristic potentials} of contact $\alpha$.
The emittance can be written as
\begin{equation}
E_{\alpha \beta }= D_{\alpha  \beta } - {\bf D}^{e}_{\alpha} {\bf u}_{\beta} \;\; .
\label{EM}
\end{equation}
To complete the calculation of the emittance, we next need a discussion
of the characteristic potentials.

%
\begin{figure}[t]
\vspace{-2cm}
\hspace{10mm}
 \epsfxsize = 90 mm
 \epsffile{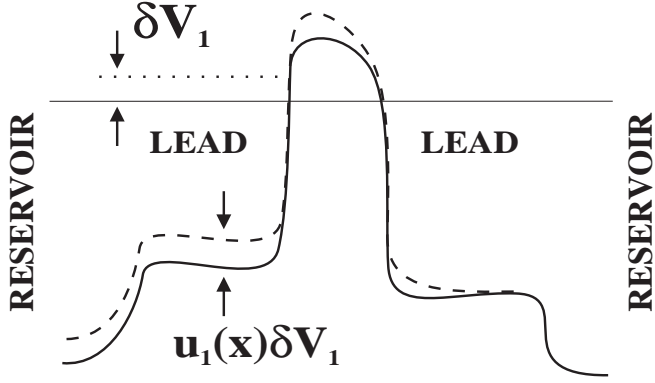}
\vspace{-5cm}
\caption{ Change (dashed) of the equilibrium potential (solid) in a two-terminal sample due to
a voltage variation $\delta V_{1}$ (dotted) in the left reservoir.}
\label{fig2}
\end{figure}

\subsection{Piled-up charge and self-consistent potential}
\label{SCP}

%
The variation of the charge $\delta q_{k}$ is the sum of a bare charge 
and a screening charge.
The bare charge can be expressed with the help of the injectivity 
\begin{eqnarray}
\delta {\bf q}^{(b)} = \sum_{\beta} {\bf D}^{i}_{\beta} \;  \delta V_{\beta} \;\;.
\label{qb1}
\end{eqnarray}
The screening charge is connected to the
electrostatic potential \cite{LEVI89a} 
via the Lindhard polarization function, which is here a matrix
${\bf \Pi}$ with elements $\Pi_{kl}$,
\begin{eqnarray}
\delta {\bf q}^{(s)} = - 
{\bf \Pi} \:  {\bf \delta U} \;\;.
\label{qs1}
\end{eqnarray}
The Lindhard matrix can be expressed in terms of the scattering states.
It is in general not a diagonal matrix, i.e. the charge response is in general nonlocal.
Non-local effects are, however, small quantum mechanical effects
which can be neglected deep inside a conductor with a sufficiently large electron 
density. In a quasiclassical treatment the Lindhard matrix 
becomes local, $\Pi_{kl} = \delta_{kl} D_{k}$.
Here $D_{k}$ is the local DOS in region $k$.\\
The total charge, $ \delta {\bf q} = \delta {\bf q}^{(b)} + \delta {\bf q}^{(s)}$,
acts now as the source of the nonequilibrium electric potential in the Poisson equation.
For our discrete model we also have to discretize the Poisson equation.
We introduce a geometrical capacitance matrix ${\bf C}$
which relates the charges  to the electrostatic potentials,
$\delta {\bf q} = {\bf C} \delta {\bf U}$.
If one uses this matrix ${\bf C}$ and expresses the electric potential 
in terms of the characteristic potentials, one can write the Poisson equation in the form
$ ({\bf C} + {\bf \Pi}) {\bf u}_{\alpha} = {\bf D}^{(i)}_{\alpha} $
which determines the characteristic potentials ${\bf u}_{\alpha}$,
\begin{equation} 
{\bf u}_{\alpha}= ({\bf \Pi}+{\bf C})^{-1}{\bf D}^{(i)}_{\alpha} \;\;.
\label{ug1}
\end{equation}
Going back to Eq. (\ref{EM}) we find for the emittance
\begin{equation}
E_{\alpha \beta }= D_{\alpha  \beta } -
{\bf D}^{(e)}_{\alpha}({\bf \Pi}+{\bf C})^{-1}{\bf D}^{(i)}_{\beta} \;\; .
\label{EMMD}
\end{equation}
We notice that the bare contribution to the emittance 
occurs with a positive sign and the Coulomb induced contribution
occurs with a negative sign. Depending on which contribution dominates
we call the emittance element {\it capacitive} ($E_{\alpha \alpha} > 0$, 
$E_{\alpha \beta } < 0$  for $\alpha \ne \beta $)
or {\it inductive-like} ($E_{\alpha \alpha} < 0$, 
$E_{\alpha \beta } > 0$  for $\alpha \ne \beta $).\\ 
The nonequilibrium charge distribution becomes 
\begin{equation}
\delta {\bf q}= {\bf C}\delta  {\bf U} =
\sum_{\beta }{\bf D}^{(i)}_{\beta}\delta V_{\beta}-{\bf \Pi}\delta { \bf U} \;\;.
\label{nonq}
\end{equation}
This determines the nonequilibrium steady-state to linear order
in the applied voltages. The nonequilibrium charge-distribution can be written
in terms of an electrochemical capacitance 
matrix $C_{\mu , k\beta}$ \cite{CHRI96c,CHRI96a,CHRI96b}
which determines the net charge variation in region $k$
in response to a potential variation at contact $\beta$. 
In vector notation, we have
\begin{equation}
\delta {\bf q} = \sum_{\beta} {\bf C}_{\mu,\beta }\delta V_{\beta}\;\; ,
\label{NECD}
\end{equation}
where
\begin{equation}
{\bf C}_{\mu,\beta } =  {\bf C} {\bf u}_{\beta} = 
{\bf C} ({\bf \Pi}+{\bf C})^{-1}{\bf D}^{(i)}_{\beta} \;\; .
\label{CMUU}
\end{equation}

We do not present here a general discussion of the charge relaxtion
term $K_{\alpha \beta }$. This requires
a dynamical screening theory, and we have results only for 
special cases. We discuss one of these special cases in Sect. \ref{PRET}
and Sect. \ref{AC-Response}.

\subsection{Weakly nonlinear dc-transport}
\label{LNDCT}
The nonequilibrium potential distribution is not only of importance 
in ac-transport but also in nonlinear transport.
Knowledge of the nonequilibrium potential distribution to first order 
in the applied voltages permits to find the nonlinear I-V-characteristic 
up to quadratic order in the voltages,
\begin{equation}
I_{\alpha} = \sum _{\beta }G_{\alpha \beta}^{(0)}V_{\beta }+
\sum _{\beta \gamma }G_{\alpha \beta \gamma}^{(0)}V_{\beta}V_{\gamma} + {\cal O} (V^{3})\;\;.
\label{secondorder}  
\end{equation} 
The coefficients $G_{\alpha \beta}^{(0)} $ and
$G_{\alpha \beta \gamma}^{(0)}$ are obtained from an expansion
of Eq. (\ref{current}) with respect to the voltages $V_{\alpha}$. 
One obtains for the linear conductance Eq. (\ref{dccond}).
An expansion of Eq. (\ref{current})
up to ${\cal O}(V^{2})$ yields 
$ G_{\alpha \beta \gamma}^{(0)} = (1/2)
\int dE\: (-\partial _{E}f)$
$\left( 2\partial _{V_{\gamma}}g_{\alpha \beta} +
e \partial _{E}g_{\alpha \beta}  \delta_{\beta \gamma}
\right)$.
Writing $ \partial _{V_{\gamma}}g_{\alpha \beta}$
in terms of the derivatives 
$dg_{\alpha \beta}/dU_{k} $ and the characteristic potentials yields 
\begin{equation}
G_{\alpha \beta \gamma}^{(0)}= \frac{1}{2} \int dE\: (-\partial _{E}f)
\sum_{k} \: \frac{dg_{\alpha \beta}}{dU(k)}
\left( 2u_{k\gamma} - \delta_{\beta \gamma}\right)\;\;.
\label{Gabc2}
\end{equation}
Expressing the characteristic potentials in terms of the injectivities we 
find that the rectification coefficient is given by
\begin{equation}
G_{\alpha \beta \gamma}^{(0)}= \frac{1}{2}\int dE (-\partial _{E}f)
\left( 2({\bf \nabla _{U}}
g_{\alpha \beta})^{t} \: ({\bf \Pi}
+ {\bf C})^{-1}{\bf D}^{(i)}_{\gamma}  +
e \delta _{\beta \gamma} \partial _{E} g_{\alpha \beta} \right)\;\;.
\label{Gabcdiscrete}
\end{equation}
For a quantum point contact, 
Eq. (\ref{Gabcdiscrete}) has been discussed in Ref. \cite{CHRI96c}.
In Sect. 3.3 we will apply this result to the resonant tunneling barrier. 


\subsection{Charge conservation, gauge invariance, and Magnetic Field Symmetry}
\label{CCGIOCR}
Since the system under consideration includes all conductors and nearby gates, the
theory must satisfy charge (and current) conservation and gauge invariance. Due to micro-reversibility
the linear response matrices must additionally satisfy the Onsager-Casimir symmetry
relations.\\
Let us first discuss charge conservation, which states
that the total charge in the sample remains constant under a bias.
This implies also current conservation, $\sum _{\alpha} I_{\alpha} =0$.
Imagine a volume $\Omega$ which encloses the entire conductor including
a portion of the reservoirs which is so large that at the place
were the surface of $\Omega$ intersects the reservoir all the
characteristic potentials are either zero or unity.
According to the law of Gauss, one concludes that the total charge remains constant.
Application of a bias voltage results only in a redistribution of the charge. 
If the conductor is poor, i.e. nearly an insulator, the contacts act
like plates of capacitors. In this case long-range fields exist which 
run from one reservoir to the other and from a reservoir to a portion of
the conductor. But if we chose the volume $\Omega$ to be large enough
then all field lines stay within this volume. Charge and current conservation imply
for the response coefficients the sum rules
\begin{equation}
\sum _{k} C_{\mu ,k \beta }=\sum _{\alpha} G_{\alpha \beta }^{(0)}=\sum _{\alpha} E_{\alpha \beta }
=\sum _{\alpha} K_{\alpha \beta }= \sum _{\alpha} G_{\alpha \beta \gamma}^{(0)}=0 \;\;.
\label{sumrules1}
\end{equation} 
Second consider the fact that only voltage differences are physically meaningful. 
Gauge invariance means that measurable quantities are invariant
under a global voltage shift $\delta V$ in the reservoirs,
$\delta V_{\alpha} \mapsto  \delta V_{\alpha} + \delta V$. A global voltage shift corresponds,
of course, only to a change of the global voltage scale. 
Consequently, the characteristic potentials satisfy \cite{BUTT93a}
\begin{equation} 
\sum_{\alpha}{\bf u}_{\alpha} =  {\bf 1} \;\;.
\label{ug1s}
\end{equation}
For the response coefficients the following sum rules hold 
\begin{equation}
\sum _{\beta} C_{\mu ,k \beta }=\sum _{\beta} G_{\alpha \beta} ^{(0)}=\sum _{\beta} E_{\alpha \beta}=\sum _{\beta} K_{\alpha \beta} =
\sum _{\beta} (G_{\alpha \beta \gamma}^{(0)}+
G_{\alpha \gamma \beta }^{(0)})=0 \;\; .
\label{sumrules2}
\end{equation} 
Note that due to charge conservation and gauge invariance,
the geometrical capacitance matrix ${\bf C}$ has  the zero mode
${\bf 1} = (1, 1, ....1)$, corresponding to a unit  
potential in each region. Thus ${\bf C}$ cannot be inverted. However, the 
Green's function $({\bf \Pi}+{\bf C})^{-1}$ which solves the Poisson equation
exists (see Eq. (\ref{ug1})). Expressing the characteristic potentials 
with the help of the Green's function and the injectivities gives
\begin{equation} 
{\bf 1} = ({\bf \Pi}+{\bf C})^{-1}{\bf D}\;\;
\label{ug2}
\end{equation}
where ${\bf D} = (D_{1}, D_{2}, ...,D_{M})$
is the vector of the local DOS.
Equation (\ref{ug2}) is the key property of the Green's function needed 
to show that our final results are charge and current conserving.
From Eq. (\ref{ug2}) it follows immediately that the sum over the elements of a row of
the Lindhard matrix is equal to the DOS in region $k$, i.e. $D_{k} = \sum_{l} \Pi_{kl}$.
Since the Lindhard matrix is symmetric, it holds also $D_{k} = \sum_{l} \Pi_{lk}$.
The above mentioned statement, that in the quasiclassical local case $\Pi _{kl} = D_{k}\delta _{kl}$
holds, is now clear.\\
In the presence of a magnetic field, the admittance matrix must additionally satisfy 
the reciprocity relations 
\begin{equation}
G_{\alpha \beta}(\omega )|_{B}
=G_{\beta \alpha}(\omega)|_{-B} \;\;
\label{OCSR}
\end{equation}
which is a consequence of time-reversal symmetry of the microscopic equations and 
the fact that we consider transmission from one 
reservoir to another \cite{BUTT93d,BUTT86a,BUTT88b}.
These Onsager-Casimir relations are only valid for the linear response coefficients,
i.e. close to equilibrium. Of course, the symmetry relations (\ref{OCSR})
hold individually for $G_{\alpha \beta}^{(0)}$, $E _{\alpha \beta}$, and $K_{\alpha \beta}$.
Note that the emittance is symmetric, $E _{\alpha \beta}=E _{\beta \alpha }$,
in the purely capacitive case.

\subsection{Frequency-Dependent Transport: Single Potential Approximation}
\label{PRET}
In this section we show that it is possible to discuss the full frequency-dependence
of the admittance, if the electric potential in the mesoscopic conductor can be approximated by
a single variable. The {\it external response} is defined as the response for
fixed electrostatic potential (i.e. the `bare' response). A general
result for the external response has been derived in Ref. \cite{BUTT93b}. If
the differences of wave vectors $k(E)$ and $k(E+\hbar \omega)$
deep in the reservoirs are neglected, 
the external response can be expressed in terms of the scattering matrix only, 
\begin{equation}
G_{\alpha \beta}^{e} (\omega ) = \frac{e}{h} 
\int dE 
Tr[{\bf 1}_{\alpha}\delta _{\alpha \beta}-{\bf s}^{\dagger}_{\alpha\beta}(E)
{\bf s}_{\alpha\beta}(E+ \hbar \omega)]
\frac{(f(E)-f(E+ \hbar \omega ))}{\hbar \omega} \;\; .
\label{eq56}
\end{equation}
The scattering matrices and the Fermi functions are here evaluated at
the equilibrium reference state.  
The real part of the ac-conductance, (\ref{eq56}) is related via the
fluctuation-dissipation theorem  to the current-current 
fluctuation spectra derived in Ref. \cite{BUTT92}. An expansion of Eq. (\ref{eq56})
to linear order in $\omega$
gives \cite{BUTT93b}
\begin{equation}
G_{\alpha \beta} ^{e}(\omega ) = G_{\alpha \beta} ^{( 0)} - i \omega {e^{2}}
\int dE \: (-df/dE)\: (dN_{\alpha \beta}/dE) \;\; ,
\label{eq57}
\end{equation}
where the $dN_{\alpha \beta}/{dE}$ are the global PDOS,
in which the functional derivative with respect to the local potential
(and the integration over the volume $\Omega$) is replaced by a (negative) 
energy derivative. Asymptotically, for a large volume $\Omega$
the global PDOS obtained from an energy derivative
is identical to the integral of the local PDOS associated with a functional derivative
with respect to the potential according to Eq. (\ref{LDOS}). However, for a finite volume, 
there are typically small differences of the order of
the Fermi wavelength divided by the size of the volume $\Omega$.
For a more detailed discussion, we refer the reader to Ref. \cite{GASP96a}.\\
In order to obtain the full admittance, we have still to find the internal
response (the `screening' part of the admittance).
A general result for the internal response is not known.
In the simple case, where the conductor is treated with a single discrete 
region ($M =1$), the external response defines, however, also the internal response.
We assume that the sample is in close proximity to a gate 
which couples capacitively to the conductor with a capacitance coefficient
$C$. The current response at contact $\alpha$ is 
$\delta I_{\alpha} (\omega) = \sum G^{e}_{\alpha \beta} (\omega ) \delta V_{\beta} +
G^{i}_{\alpha} (\omega ) \delta U_{1} $
where $G^{i}_{\alpha} (\omega )$ is the (unknown) internal response 
of the conductor generated by the oscillating electrostatic potential $\delta U_{1}$.
The current induced into the gate is 
$\delta I_{g} (\omega) = -i \omega C(\delta V_{g} - \delta U _{1})$.
Now we can determine  $G^{i}_{\alpha} (\omega )$ from the requirement
of gauge invariance. In particular, if all potentials are shifted  
by $ - \delta U _{1}$, it follows immediately that 
$G^{i}_{\alpha} (\omega ) = - \sum_{\beta} G^{e}_{\alpha \beta} (\omega )$. 
Using current conservation, $\sum _{\alpha} \delta I_{\alpha}+\delta I _{g}=0$,
we find for the conductances \cite{BUTT93b}
of the interacting system ($\alpha ,\beta, \gamma, \delta \neq g $),  
\begin{eqnarray}
G^{I}_{\alpha\beta} (\omega)
& = & G^{e}_{\alpha \beta} (\omega )- 
\frac{(i/\omega C) \sum_{\gamma} G^{e}_{\alpha \gamma} (\omega )
\sum_{\delta} G^{e}_{\delta \beta} (\omega )}
{1+(i/\omega C) \sum_{\gamma\delta} G^{e}_{\gamma\delta} (\omega )} , 
\label{eq61} \\
G^{I}_{g\beta} (\omega)
& = & - \frac{\sum_{\delta} G^{e}_{\delta \beta} (\omega )}
{1+(i/\omega C) \sum_{\gamma\delta} G^{e}_{\gamma\delta} (\omega )}  , 
\label{eq62} \\
G^{I}_{\alpha g} (\omega)
& = & - \frac{\sum_{\gamma} G^{e}_{\alpha \gamma} (\omega )}
{1+(i/\omega C) \sum_{\gamma\delta} G^{e}_{\gamma\delta} (\omega )}  , 
\label{eq63} \\
G^{I}_{g g} (\omega)
& = & - \frac{\sum_{\gamma \delta} G^{e}_{\gamma \delta} (\omega )}
{1+(i/\omega C) \sum_{\gamma\delta} G^{e}_{\gamma\delta} (\omega )}  . 
\label{eq64}
\end{eqnarray}  
The single-potential approximation provides a reasonable
description only if the conductor has a well-defined interior region,
which might be described by a single uniform potential.
Furthermore, it is assumed that the only relevant capacitance is that 
of the sample and the gate.  
Examples for which this approach is 
reasonable are quantum wells or quantum dots or cavities
for which the Coulomb interaction is so weak that 
single electron effects can be neglected. 
In Section \ref{CDOAC} we discuss the case of a double barrier with a 
well which is capacitively coupled to a gate. 

%
%
%

\section{Examples}
\label{examples}

%
%
In a first example we discuss the emittance of 
a wire which contains a single barrier. The wire 
is capacitively coupled to a gate. 
As a second example we discuss the capacitance and emittance 
of a quantum point contact formed with the help of gates \cite{CHRI96b}.
It turns out that steps in the capacitance and
the emittance occur in synchronism with the well-known conductance steps. 
As a third example, we treat the
resonant tunneling barrier close to a resonance, for which we calculate
the nonlinear current-voltage characteristic and the frequency-dependent conductance 
\cite{CHRI96c} with the single-potential approximation. For all examples, we use
explicit expressions for the Lindard function (matrix)
${\bf \Pi }$ given within the local Thomas-Fermi approximation, where ${\bf \Pi }$
is diagonal with elements determined by the local DOS.
In all examples we consider finally the zero-temperature limit.

\subsection{Admittance of a wire with a Barrier}
\label{WI}
\begin{figure}[t]
\hspace{25mm}
 \epsfxsize = 60 mm
 \epsffile{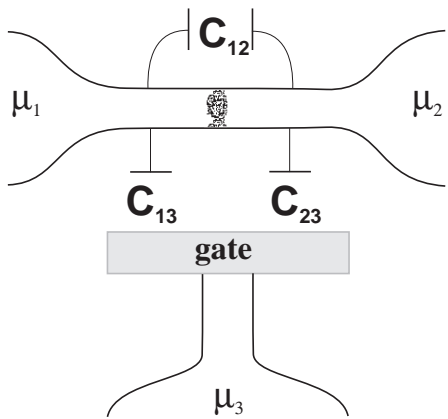}
 \vspace{-2.5cm}
\caption{ Quantum wire containing an impurity and coupled to a gate.}
\label{wirefig}
\end{figure}

Our first example is a mesoscopic wire which connects two contacts
and is capacitively coupled to a nearby gate (see Fig. \ref{wirefig}).
The wire is assumed to be perfect with a uniform potential along its 
main direction, except for a single scatterer, a barrier, or an impurity.
The discussion is carried out in the semiclassical limit where
Friedel oscillations are neglected. We take the potential
of the impurity to be of short-range and assume that the potential 
to left and right of the scatter can be assumed to be uniform.
We assume here that the gate couples to the wire only and neglect 
scattering at the interface of the wire and the reservoir.
Furthermore, to simplify the discussion we assume in this 
section that the capacitance 
between the charges to the left and the right of the barrier can be 
neglected (i.e. vanishingly small $C_{12}$).
Consider first a perfect wire ($T = 1$) of length $L$.
The gate couples with a geometrical capacitance $c$ per length. 
%
The DOS per length for a one-dimensional wire is 
\begin{eqnarray}
dn/dE = 2/hv 
\label{MB2}
\end{eqnarray}
where $v = (2/m)^{1/2} (E - eU)^{1/2}$ is the velocity of carriers with energy $E$.
If the wire is filled up to the electrochemical potential $\mu$ of the reservoirs
the total number $n$ of states per length is 
\begin{eqnarray}
n(\mu) = \int_{eU}^{\mu} dE (dn/dE)= e^{-1}(2/\pi^{2} a_{B})^{1/2}(\mu - eU)^{1/2} .
\label{MB3}
\end{eqnarray}
Here we have introduced the Bohr radius $a_{B} = \hbar^{2}/me^{2}$.
To determine this density self-consistently we consider the Poisson equation.
The difference of the electron charge density $en$ and the ionic 
background charge density $-en^{+}$
is equal to the capacitive charge,
\begin{eqnarray}
en(\mu) - en^{+} = c(U - U_{g}) \;\;,
\label{MB4}
\end{eqnarray}
where $U_{g}$ is the electrostatic potential of the gate.
Similarly, near its surface, the deviation of the electronic gate charge 
away from charge neutrality is equal to $- c(U - U_{g})$. 
Now we take the gate to be a macroscopic conductor.
Therefore, the electrostatic potential and the electrochemical potential
are everywhere related by $\mu_{g} = E_{Fg}+eU_{g}$,
where $E_{Fg}$ is the chemical potential (Fermi energy) of the gate. 
We eliminate the gate potential $U_{g}$ in Eq. (\ref{MB4})
and find
\begin{eqnarray}
en(\mu) =(c/e) (e^{2}n^{+}/c + E_{Fg} - \mu_{g} + eU) \;\; .
\label{MB5}
\end{eqnarray}
Measuring energies from the band bottom of the one-dimensional 
wire and setting $eU = 0$ gives the electrochemical potential as a function
of the gate potential,
\begin{eqnarray}
\mu =(c/e)^{2} (\pi^{2} a_{B}/2) (e^{2}n^{+}/c + E_{Fg} - \mu_{g})^{2} \;\; .
\label{MB6}
\end{eqnarray}
In the limit $c \to 0$ the electrochemical potential of the wire 
is simply determined by charge neutrality,
$\mu = $ $(\pi^{2} a_{B}/2)e^{2}$ $(n^{+})^{2}$. 
For $\mu_{g}= \mu_{0} \equiv e^{2}n^{+}/c + E_{Fg}$ the electrochemical potential
vanishes and the wire is empty for  $\mu_{g} \geq \mu_{0}$.
The DOS of left moving carriers, $dn_{l}/dE = 1/hv$,
is directly related to the phase accumulated by the 
electrons traversing the wire, $Ldn_{l}/dE = L/hv = (1/\pi) d\phi /dE$.
The phase accumulated by a traversing electron (at the Fermi energy) is thus given by
\begin{eqnarray} 
\phi = \pi L (c/e^{2}) (e^{2}n^{+}/c + E_{Fg} - \mu_{g}) \;\; .
\label{MB7}
\end{eqnarray}
The scattering matrix element for the transmitting channel is given by $s_{21} =  \exp(i\phi)$ 
which shows that the scattering matrix and thermodynamics are intimately 
related. At the Fermi energy of the wire, the DOS is given by
\begin{eqnarray} 
dn/dE =  (1/2\pi^{2} a_{B})(1/c) (e^{2}n^{+}/c + E_{Fg} - \mu_{g})^{-1} .
\label{MB9}
\end{eqnarray}
The true electron density in a narrow energy interval at the Fermi energy
is found by considering a small variation of the electrochemical potentials 
$\delta \mu$ and $\delta \mu_{g}$ and the electric potential $\delta U$
away from a reference state ($\mu^{(eq)}$, $\mu_{g}^{(eq)}$, $U^{(eq)}$)
which obeys Eq. (\ref{MB5}). Variation of this equation gives 
\begin{eqnarray}
\delta q =  e(dn/dE)(\delta \mu -e\delta U) = c (\delta U - \delta \mu_{g}/e).
\label{MBa1}
\end{eqnarray}
Solving for $\delta U$ gives for the charge  
$\delta q =  c_{g} (\delta \mu /e  - \delta \mu_{g}/e) $
with an electrochemical wire-to-gate capacitance 
\begin{eqnarray} 
c^{-1}_{g} = c^{-1} + (e^{2} (dn/dE))^{-1} \;\;.
\label{MBwg}
\end{eqnarray}
Here the DOS appears in series with the geometrical capacitance
It depends on the reference state and is given by Eq. (\ref{MB9}) 
with $\mu_g = \mu^{eq}_{g}.$\\
Next, we discuss a wire containing a symmetric barrier.
At equilibrium, and if we can treat the potential created by the 
barrier as short-range, the DOS to the left and right
of the barrier remains unchanged.  
Hence the consideration given above for the electrostatic equilibrium potential 
also characterizes the wire to the left and right of the barrier.
Similarly, for a long wire the gate-to-wire capacitance  
is essentially unaffected by the presence of the barrier. 
In the case of transport, on the other hand, 
the chemical potentials of the reservoirs connected to the wire 
differ by a small amount implying a variation of the charge distribution. We
introduce two regions $\Omega_{1}$ and  $\Omega_{2}$ of size $L/2$ to the left
and the right of the barrier, respectively. 
With the help of Fig. \ref{fig1}b, the semiclassical local PDOS, $D_{\alpha k \beta}$,
can be constructed with simple arguments.
For example, $D_{211}$
is given by the transmission probability times the DOS of $\Omega _{1}$
associated with carriers with positive velocity, hence $D_{211}=T D_{1}/2$,
with $D_{1}=D_{2}= D/2 $ due to symmetry. To determine $D_{212} $, we note
that in the semiclassical limit considered here, there
are no states in $\Omega _{1}$ associated
with scattering from contact $2$ back to contact $2$, hence it holds
$D_{212} =0$. With similar arguments one finds for the semi-classical PDOS
\begin{equation}
D_{\alpha k \beta} =
D_{k} \left( T/2 + \delta _{\alpha \beta}(R\: \delta _{\alpha
k}-T/2)\:\right) \;\;, \;\;\; {\rm if }\;\; \alpha , \beta \neq 3 \;\;.
\label{lpdosqpc}
\end{equation} 
Form Eq. (\ref{lpdosqpc}) we obtain for the emissivities and injectivities 
$D^{e}_{11} = D^{i}_{11}=(1/4)(1+R) D$ and 
$D^{e}_{12} = D^{i}_{12}=(1/4)TD$. The integrated geometrical capacitance 
is denoted by $C = cL$. With these quantities the Poisson equation reads 
\begin{eqnarray}
(D/2)(1+R) \delta V_{1} + (TD/2) \delta V_{2} - D \delta U_{1}
& = & C (\delta U_{1} - \delta U_{3})
\nonumber \\
(TD/2) \delta V_{1} + (1+R) (D/2) \delta V_{2} - D \delta U_{2}
& = & C (\delta U_{2} - \delta U_{3})
\label{MB11b}
\end{eqnarray}
These equations can also be derived by noticing that the variations 
$\delta F_{k}$ of the
quasi-Fermi levels \cite{BUTT85a} in the regions $k=1$ and $2$ are given by
$\delta F_{1}=((1+R)\delta V_{1} + T\delta V_{2} )/2$ and
$\delta F_{2}=((1+R)\delta V_{2} + T\delta V_{1} )/2$, respectively.  
For a local Lindhard function, 
the charge in region $k$ 
is then $\delta q _{k}=D_{k}(\delta F_{k}-\delta U_{k})=(C/2)(\delta U_{k}-\delta U_{3})$.\\
For the charge neutral case, $C = 0$, Eqs. (\ref{MB11b})
are familiar from Landauer's discussion of the potential drop across a 
barrier \cite{LAND70a,LAND57a,BUTT85a}. 
We assume again that the gate is a massive conductor for 
which we have $\mu_{3} = E_{Fg} +eU_{3}$ everywhere. 
We use  Eqs. (\ref{MB11b})  to evaluate the characteristic potentials and get
\begin{eqnarray}
u_{11}= \frac{1}{2} \frac{(1+R)D}{C+D} , 
& u_{12} =  \displaystyle\frac{1}{2} \frac{TD}{C+D} , 
& u_{13} = \frac{C}{C+D}\;\;.
\label{MB15}
\end{eqnarray}
Due to symmetry, it holds  
$u_{22} = u_{11}$, $u_{21} = u_{12}$ and $u_{23} = u_{13}$.
Defining $\Delta V\equiv \delta V_{1}-\delta V_{2}$,
the potential drop across the barrier can be written in the form
\begin{eqnarray}
\Delta U  \equiv \delta U_{1} - \delta U_{2}= (C_{g}/C) R \Delta V
\label{MB16}
\end{eqnarray}
where $C_{g} = Lc_{g}$ is the electro-chemical
gate-to-wire capacitance given by Eq. (\ref{MBwg}).
The voltage difference $\Delta U$
is proportional to the reflection probability $R$ and the 
chemical potential difference of
the reservoirs. The total charges to the left or to the right of the barrier are 
the sum of the injected charges and the induced charges 
\begin{eqnarray}
\delta q_{k} =\sum_{\beta}  (D^{i}_{ k \beta} - D_{k}u_{k \beta}) \delta V_{\beta} \;\;.
\label{MB17}
\end{eqnarray}
We find, e.g., for $\delta q_{1}$,
\begin{eqnarray}
\delta q_{1} = C_{g}  \{ (1/2)(1+R) \delta V_{1} + (T/2) \delta V_{2} - \delta V_{3}\}.
\label{MB18}
\end{eqnarray}
The difference in the charge density on the left and the right hand side becomes
\begin{eqnarray}
\Delta q \equiv \delta q_{1} -\delta q_{2} = C_{g} R \Delta V.
\label{MB19}
\end{eqnarray}
In the limit of vanishing gate-to-wire capacitance,
the charge difference vanishes also. Away from the barrier the wire is 
charge neutral. In this charge-neutral limit the potential difference is determined
by the reflection coefficient only, $\delta U= R \Delta V$.\\
From  Eq. (\ref{MB18}) we find
the electrochemical capacitance coefficients $C_{\mu , k \alpha}$.
We write this capacitance matrix in the form
\begin{equation}
{\bf C}_{\mu } = 
\left(
\begin{array}{ccc}
 C_{\mu }  &C_{g} - C_{\mu }  & - C_{g} \\
C_{g} - C_{\mu } & C_{\mu } & - C_{g} \\
- C_{g} & -C_{g}  & 2 C_{g} 
\end{array}
\right) \;\; ,
\label{cmumat}
\end{equation}
where 
\begin{eqnarray} 
C_{\mu} = (1/2)(1+R) C_{g} .
\label{MB21}
\end{eqnarray}
It follows $C_{\mu ,12} = (1/2) T C_{g}$. 
Using the characteristic potentials and the PDOS
given by Eq. (\ref{lpdosqpc}) we can calculate the emittance, 
\begin{eqnarray}
E_{11} & = & E_{22} = (C_{g}/C) (RC- DT^{2}/4) \nonumber \\
E_{12} & = & E_{21} = (C_{g}/C) (TC+ DT^{2}/4)\nonumber \\
E_{13} & = & E_{31} = E_{23}  =   E_{32} = - E_{33}/2 = - C_{g}
\label{MB1}
\end{eqnarray}
If the transmission probability vanishes, the emittance matrix
is purely capacitive. In the case where the geometric gate capacitance $C$ tends to zero, 
the electrochemical gate capacitance and the electrochemical capacitance across
the barrier vanish $C_{g} = C_{\mu} = 0$ but  the ratio $C_{g}/C$ tends to $1$. 
The wire is then charge neutral and the charge imbalance $\Delta q$ 
vanishes. In the single-channel case considered here,
the voltage drop is $\Delta U = R\Delta V$.
In this case the wire responds inductively for all values of the transmission probability
$E = - T^{2} D/4 $. In the absence of a barrier ($R = 0$) the wire has also inductive 
emittances $E = E_{11} = E_{22} = - E_{12} = - E_{21} = - D/4 $.
The results of this section are summarized in the rightmost column of table 1.

%
\begin{table}
\begin{center}
\begin{minipage}{20.5pc}
\caption[Gate capacitance and capacitance, emittance, polarization, and voltage drop of
 the quantum point contact with a grounded gate for limiting cases ]{Gate
 capacitance and capacitance, emittance, polarization, and voltage drop of
 the quantum point contact with a grounded gate for limiting cases}
\begin{tabular}{|l||c|c|c|} \hline
             & $C \to 0 $                 &$ C \to \infty $       &$ C_{0} \to 0 $ \\ 
             & $  C_{0} = $ const.        &$ C =   $ const.       &                \\ \hline\hline
$C_{g} $   & $0$            & $D/2$             & $1/((D/2)^{-1}+C^{-1})$      \\ \hline
$C_{\mu} $  & $R/(C_{0}^{-1}+D^{-1})$            
                              & $(D/4)(1+R)$      & $(1+R)C_{g}/2  $              \\ \hline
$E_{11}$    & $RC_{\mu}-DT^{2}/4$            
                              & $RD/2$            & $(C_{g}/ C)(RC -DT^{2}/4)$  \\ \hline
$\Delta q / \Delta V $   
             & $2C_{\mu}$     & $RD/2$             & $RC_{g}$        \\ \hline
$ \Delta U / \Delta V$        
             &$C_{\mu }/C_{0}$&     $0$             & $R C_{g}/C$       \\ \hline
\end{tabular} 
\end{minipage}
\end{center}
\label{table1}
\end{table}

%

\subsection{Emittance of a quantum point contact in the presence of gates}
\label{QPC}

\begin{figure}[t]
\hspace{1mm}
 \epsfxsize = 50 mm
 \epsffile{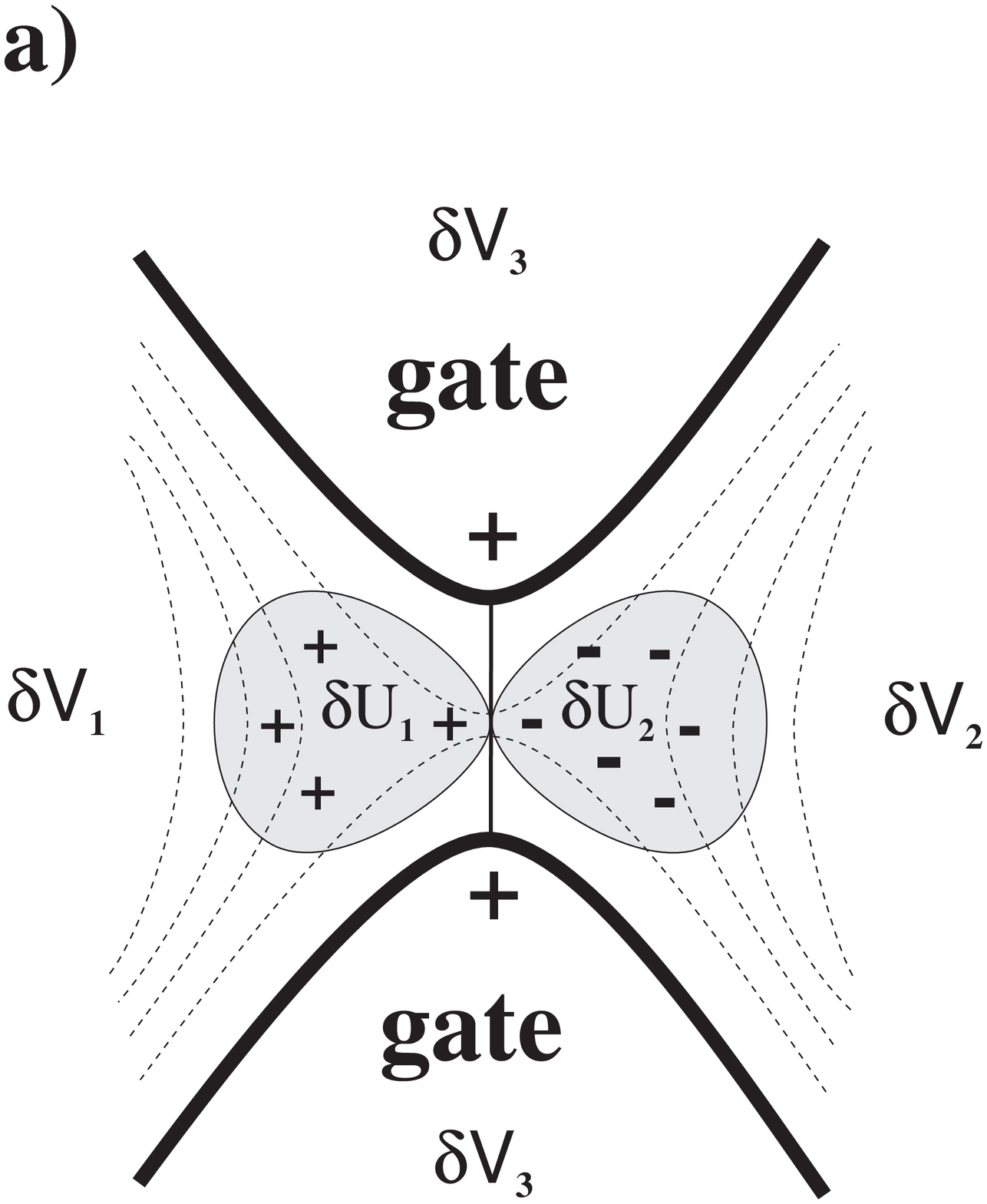}
\hspace{5mm}
 \epsfxsize = 55 mm
 \epsffile{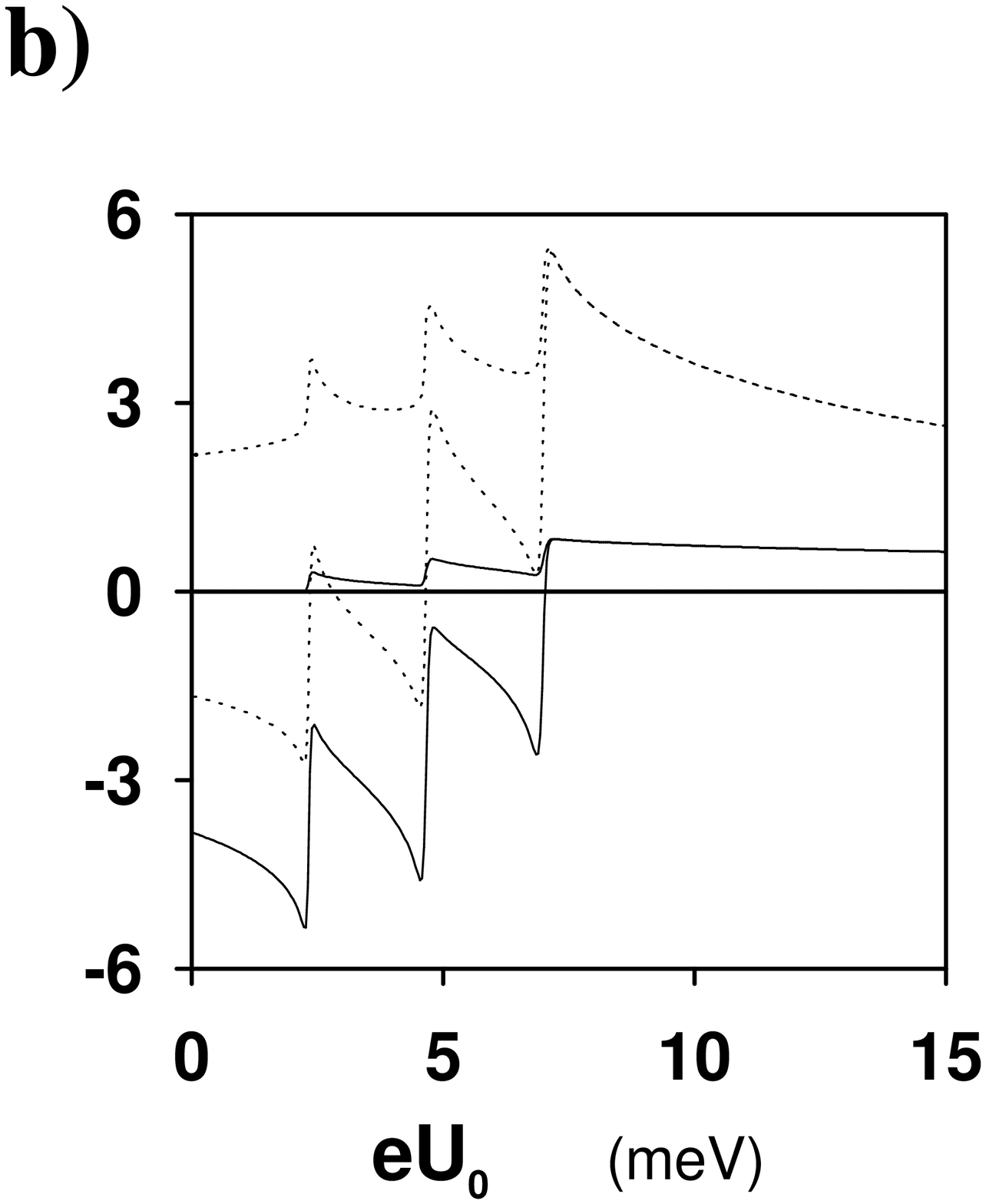}
 \vspace{-0.5cm}
\caption{ a) Quantum point contact with gates. The regions $\Omega _{1,2}$
at the left and right of the point contact can be charged. b) Capacitance $C_{\mu }$
and emittance $E_{11}$ of the QPC. Capacitances and emittances are in units of 
femto farads.}
\label{fig5}
\end{figure}

A quantum point contact (QPC) is a small constriction in a two-dimensional electron gas
which allows the transmission of only a few conducting channels \cite{VANW88a,WHAR88a}.
We consider a symmetric QPC with two gates as shown in Fig. \ref{fig5}a. and ask for its
capacitance and low-frequency admittance. To discretize the QPC,
we define to the left and to the right of the constriction
two regions $\Omega _{1}$ and $\Omega _{2}$ with sizes of the order of the screening length.
If only a few channels are open, 
the equilibrium potential of the constriction has the form of a saddle \cite{BUTT90c}.
Away from the saddle towards the two-dimensional electron gas, the potential has the 
form of a valley which rapidly deepens and widens.
In contrast to the previous example, $\Omega _{1}$ and $\Omega _{2}$
are now regions immediatley to the 
left and right of the saddle, where the equilibrium potential and the equilibrium density 
are not uniform. \\  
Consider a symmetric QPC. The two gates are taken to be at the same voltage $V_{3}$.
Thus in effect, the two gates act like a single gate.
As in the previous section, the gate will be treated as a macroscopic conductor. 
For the QPC contact in the presence of the gates 
the charge distribution is not dipolar. It is the sum of the charges
in $\Omega_{1}$, $\Omega_{2}$ and at the gates which vanishes,
$\delta q_{1}+\delta q_{2}+\delta q_{3} = 0$.
%
Assume a single open channel. Due to the symmetry of the problem,
the geometric capacitance matrix can be written in the form
\begin{equation}
{\bf C} = 
\left(
\begin{array}{ccc}
C_{0}+ C& -C_{0} & -C \\
-C_{0} & C_{0}+C & -C\\
-C & -C& 2C 
\end{array}
\right) 
\;\;\;,
\label{cgeo}
\end{equation}
where $C_{0}$ is the geometric capacitance between $\Omega_{1}$ 
and $\Omega_{2}$ and where $C$ is the geometric capacitance
between these regions and the gates.
The electrochemical capacitance has again the form (\ref{cmumat}),
with $C_{\mu }$ equal to the electrochemical capacitance across the QPC and 
with $C_{g}$ being the electrochemical gate capacitance. 
With the help of Eq. (\ref{CMUU}) and  the characteristic potentials
$u_{k \beta}=(D_{k\beta}-C_{\mu , k \beta })/D_{k}$ one finds 
\begin{eqnarray}
C_{g} & = & \frac{1}{C^{-1}+ (D/2)^{-1} } 
\label{Cgqpc} \\
C_{\mu} & = & \frac{ RC_{0}+ (C/2) (1+R) + 2C_{0}C_{g}D^{-1} }{1+
2(2C_{0}+ C)D^{-1}} \;\; .
\label{Cmqpc}
\end{eqnarray}
The emittance elements (\ref{EMMD}) become
\begin{eqnarray}
E_{11} & = & E_{22} = R C_{\mu} -DT^{2}/4 + C_{g} T/2 \nonumber \\
E_{12} & = &  E_{21} = C_{g}  - E_{11} \nonumber \\
E_{13} & = &  E_{31} = E_{23}  =   E_{32} = - E_{33}/2 = - C_{g} 
\label{E11}
\end{eqnarray}
with $D=2D_{1}$. The charge difference $\Delta q$ and the electrostatic voltage
drop $\Delta U = \delta U_{1} -\delta U_{2}$ across the QPC are, respectively, given by
\begin{eqnarray}  
\frac{\Delta q }{\Delta V} & = &
\frac{DR(2C_{0}+ C)}{D+ 2(2C_{0}+C)} \;\;,
\label{C3} \\
\frac{\Delta U}{\Delta V} & = & \frac{DR}{D+ 2(2C_{0}+C)} \;\;.
\label{C4}
\end{eqnarray}
The discussion in the previous section
is a limiting case of the general results given here.
In the limit $C_{0} \to 0$ we obtain the results discussed for 
the quantum wire with a barrier
collected in column three of table 1. 
In the limit $C \to 0$ overall charging of the QPC is prohibitive.
The charge distribution is dipolar and the results of column $1$
of table 1 apply. In the limit $C \to \infty$ (corresponding to a geometrical 
capacitance of the QPC to the gates which is 
much larger than the capacitance across the QPC) 
an electrostatic voltage difference across the QPC does not develop, 
even so a charge imbalance exists. The results for this limiting case
are collected in the middle column of table 1.\\
To generalize these results to the many channel case, we mention 
that for the potential considered the contributions 
of each individual channel just add. Each occupied subband $j$ contributes 
with a transmission probability $T^{(j)}$ to the total transmission
function $ {\cal T}$ $(=T_{11}) =$ $ \sum _{j} T^{(j)}$. The dc-conductance (\ref{dccond}) is then
given by $G_{11}^{(0)}=G_{22}^{(0)}=-G_{12}^{(0)}=-G_{21}^{(0)}=(2e^{2}/h){\cal T}$. Clearly,
$G_{\alpha \beta}^{(0)}$ vanishes whenever one of the indices corresponds to the gate index $3$.
Similarly
the local PDOS (\ref{lpdosqpc}) $D_{k}^{(j)}$ of each individual channel simply add.
Hence, the local DOS becomes  $D_{k}=\sum _{j}D_{k}^{(j)}$,
and $T=1-R$ is now an average transmission probability
defined by $T\equiv T_{k}=D_{k}^{-1}\sum _{j}D_{k}^{(j)}T^{(j)}$ where ($k=1,2$).
Note that the average transmission probability
$T$ ($ \neq {\cal T}$) has nothing to do with the dc-conductance.
If only a few channels are open the potential of the QPC has the from 
of a saddle point \cite{BUTT90c}.
As a specific example we consider a quadratic potential
$U(x)=U_{0}(b^{2}-x^{2})/b^{2}$ if $|x|\leq b$,
and $U(x)=0$ if $b<|x|\leq l$. The transmission probabilities $T^{(j)}$ and
the PDOS can be calculated analytically
from a semiclassical analysis \cite{MILL53a}. For simplicity, we assume a
constant electrostatic capacitance $C_{0}= 1\: fF$ between $\Omega
_{1}$ and $\Omega _{2}$ and a fixed number
of occupied channels in these regions. Both the height $U_{0}$
and the spatial variation of the potential should in principle be found
by minimizing the grandcanonical potential for the system \cite{SHKL}.
Here we consider $U_{0}$ as an independent control parameter. 
We assume that no additional
channels enter into the regions $\Omega _{k}$ during the variation of
$U_{0}$. In Fig. \ref{fig5}b. we show the results
for a constriction with $b=500\: nm$, $l = 550 \: nm$, and
with three equidistant channels separated by $E_{F}/3=7/3 \: meV$.
The dotted curve represents the transmission function which determines
the dc conductance. At each step a channel is closed.
The dashed and solid curves correspond to a gate with coupling
$C = 0$ and $C = C_{0}$, respectively. The curves represent
the capacitance $C_{\mu}$ and the emittance $E_{11}$. For the
two-terminal QPC ($C =0$) the  capacitance vanishes and the emittance is 
negative for small $U_{0}$ where all channels are open. 
At each conductance step, the capacitance and the emittance
increase and eventually merge when all channels are closed.
Due to a weak logarithmic divergence of the WKB density of
states at particle energies $E=eU_{0}$ (where WKB is
not appropriate), the emittance shows steep edges
between the steps. In the presence of the gate ($C \neq 0$),
the curves are shifted upwards due to a capacitive contribution
of the gate. 

\subsection{The resonant tunneling barrier}
\label{CDOAC}
\begin{figure}[t]
\hspace{1mm}
 \epsfxsize = 50 mm
 \epsffile{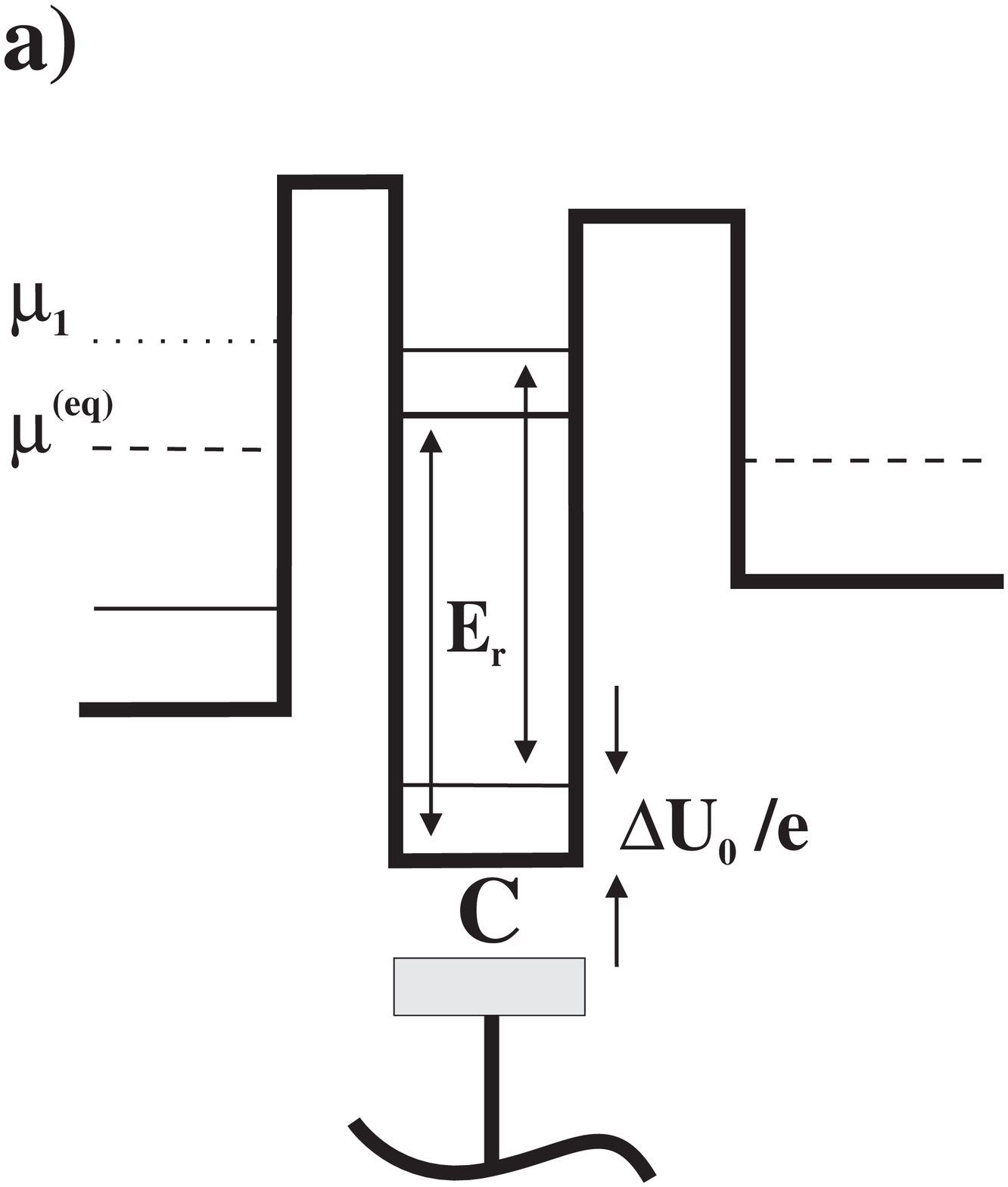}
\hspace{5mm}
 \epsfxsize = 55 mm
 \epsffile{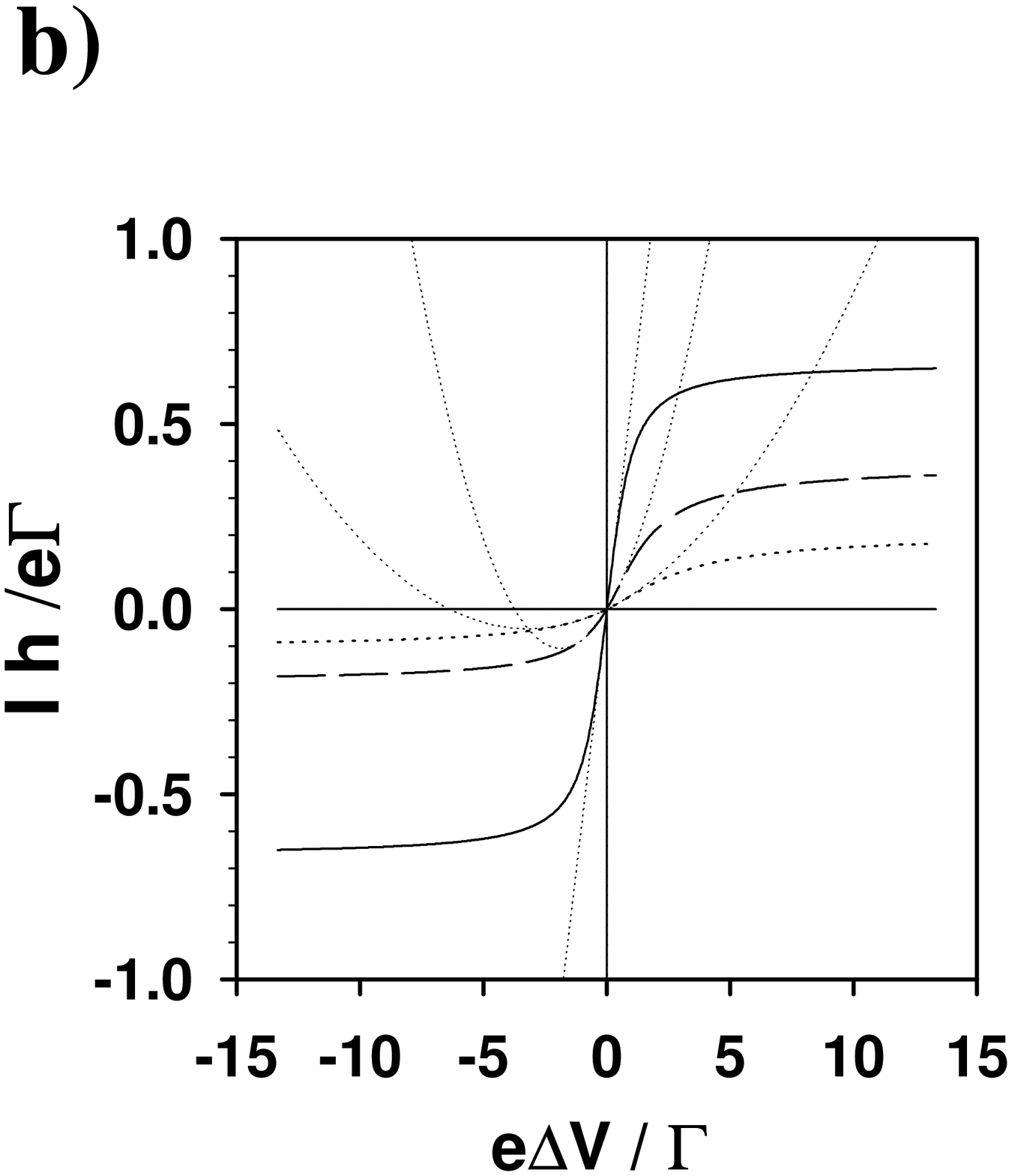}
 \vspace{-0.5cm}
\caption{ a) Resonant tunneling barrier with a single level and biased by
voltages $V_{1}=(\mu _{1}-\mu ^{(eq)})/e$ and
$V_{2}=0$. The well may be coupled capacitively with a capacitance $C$ to
a gate (dark region).
b) Nonlinear current-voltage characteristic
with $\Gamma _{1}= 0.5$ meV, $\Gamma _{2} = 1 $ meV, $\Delta E = 0$
(solid), $\Delta E = -\Gamma $ (dashed), $\Delta E = -2\Gamma $ (dotted).
Thin dotted lines indicate the quadratic approximation.}
\label{rtbfig1}
\end{figure}

In the framework of the single-potential approximation introduced in Sect. \ref{PRET}
we discuss now the fully nonlinear $I$-$V$ characteristic and
the fully frequency-dependent admittance of a resonant tunneling
barrier (RTB) with a single resonant level.
For a comparison of the results of our Hartree-like discussion
(which to be realistic needs to be extended to a many level, many channel RTB) 
with theories that treat single electron effects, 
we refer the reader to the works by 
Bruder and Sch\"{o}ller \cite{BRUD94a}, Hettler and Sch\"{o}ller \cite{SCHO95a},
Stafford and Wingreen \cite{STAF96a} and Stafford \cite{STAF96b}. 
Consider the RTB sketched in Fig. \ref{rtbfig1}a, where two reservoirs
are coupled by a double barrier structure. The well between the barriers
is considered to be the single relevant region. Additionally, 
the well may couple electrically with a capacitance $C$ to an external gate.
We label quantities associated with the well with an index $0$. 
The band bottom of the
well between the barriers and the energy of the resonant
level are denoted by $eU_{0}$ and by $E_{0}=eU_{0}+E_{r}$, respectively. 
The electrochemical potentials of the particles which are scattered at the 
RTB are assumed
to be close to the resonant level, such that the energy dependence
of the scattering properties is described by a Breit-Wigner resonance
with a width $\Gamma $. The asymmetry of the barrier is denoted by $\Delta \Gamma =
\Gamma _{1}-\Gamma _{2}$, where $\Gamma _{\alpha}/ \hbar$ is
the escape rate of a particle trapped in the well through the barrier $\alpha $ .  
The scattering matrix elements have then a pole associated with a denominator
$\Delta (E)= E-E_{0}+i\Gamma /2$,
\begin{equation}
s_{\alpha \beta }= e^{i(\varphi  _{\alpha}+\varphi _{\beta})}\left(\delta _{\alpha \beta}-
i\frac{\sqrt{\Gamma _{\alpha}\Gamma_{\beta}}}{\Delta (E)} \right) \;\;.
\label{rtb1}
\end{equation}
Here, the $\varphi _{\alpha}$ are arbitrary phases. We assume that the only energy dependence
occurs in the resonant denominator, while the $\Gamma _{\alpha}$ and the $\varphi _{\alpha}$
are energy independent. One finds a transmission probability 
\begin{equation}
T(E) = 1-R(E) = \frac{\Gamma _{1}\Gamma _{2}}{|\Delta (E)|^{2}} 
\label{rtb2}
\end{equation}
with a maximum value $T_{max}= 4\Gamma _{1} \Gamma _{2}/\Gamma ^{2}$.
The injectivities and emissivities are equal to each other
due to the absence of a magnetic field and are
\begin{equation}
D_{0\alpha }(E) = D_{\alpha 0}(E) =
\frac{e^{2}}{2\pi }\: \frac{\Gamma _{\alpha} }{|\Delta (E)|^{2}} \;\; .
\label{rtb3}
\end{equation}
Summing up injectivities (or emissivities) yields the total DOS in the well
\begin{equation}
D_{0}= e^{2}\Gamma /2\pi |\Delta (E)|^{2}\;\; .
\label{dosinwell}
\end{equation} 
Again, the (P)DOS are here represented in units of a capacitance.  
%
%
%
%
%
%
%

\subsubsection{Nonlinear transport}
Let us discuss the nonlinear $I$-$V$ characteristic \cite{CHRI96c}
for the charge neutral case ($C=0$). 
Note that only the effect of the resonant level is considered and that transport is still
phase coherent. First, we determine the self-consistent potential.
The relation between the
electrostatic potential shift $\Delta U_{0}\equiv U_{0}-U_{0}^{(eq)}$ and the voltage shifts
$V_{\alpha} = (\mu _{\alpha}-\mu ^{(eq)})/e $
is obtained from the charge-neutrality condition 
\begin{equation}
\int _{-\infty}^{\mu_{1}} D_{01}(E,U_{0})\: dE+
\int _{-\infty}^{\mu_{2}} D_{02}(E,U_{0}) \: dE
-\int _{-\infty}^{\mu ^{(eq)}} D_{0}(E,U_{0}^{(eq)}) \: dE \equiv 0
\;\;,
\label{cneutr}
\end{equation}
which can be integrated analytically. The result can be written
as a function which determines implicitly $W\equiv \Delta U_{0}-(V_{1}+V_{2})/2$
as a function of $\Delta V$ only,
\begin{eqnarray}
\Gamma \: {\rm arctan} [\frac{\Delta E}{\Gamma /2}] & = & \Gamma _{1} \:{\rm arctan}[\frac{ \Delta E - e (W-\Delta V/2)}{\Gamma /2} ]
\nonumber \\
& + & 
\Gamma _{2}\:{\rm arctan}[\frac{\Delta E - e (W+\Delta V/2)}{\Gamma /2}]   \;\;.
\label{cneutr2}
\end{eqnarray}
Equation (\ref{current}) yields for the current $I\equiv I_{1}=-I_{2}$
\begin{eqnarray}
I & = & \frac{2e}{h} \frac{\Gamma _{1}\Gamma _{2}}{\Gamma}
\biggl( {\rm arctan}[\frac{ \Delta E - e (W-\Delta V/2)}{\Gamma /2} ]
\nonumber \\
 & - &
{\rm arctan}[\frac{\Delta E - e (W+\Delta V/2)}{\Gamma /2}]   \biggr) \;\;,
\label{IVRTB}
\end{eqnarray} 
where $\Delta E = \mu ^{(eq)}-(eU_{0}^{(eq)}+E_{r})$ is the equilibrium
distance between the Fermi energy and the resonance \cite{CHRI96c}.
Without taking into account the self-consistent shift $\Delta U_{0}$ 
one would get a wrong result which is not gauge invariant.
The current given by Eq. (\ref{IVRTB}) saturates at a maximum value
proportional to $\pi /2 - {\rm arctan}(2\Delta E /\Gamma)$.
The conduction is optimal for $\Delta E =0$ and
$\Gamma _{1}=\Gamma _{2}$ when $I= (e/h)\Gamma {\rm arctan }(e\Delta
V/\Gamma)$. In Fig. \ref{rtbfig1}b we have plotted the characteristic
for an asymmetry $\Delta \Gamma /\Gamma = -1/3$ and
for various values of $\Delta E$.  
Due to the complete screening, the resonant level floates up or down
to keep the charge in the well constant. An expansion of the current
yields \cite{CHRI96c} $G_{111}^{(0)}=- (e^{3}/h)(\Delta \Gamma /2\Gamma)\partial _{E}T$
(thin dotted lines in Fig. \ref{rtbfig1}b). 
The case of incomplete screening can similarly be treated with 
our approach. At large voltages, the 
resonance can then eventually fall below the conductance band bottom
of the injecting reservoir as is known from semiconductor
double-barrier structures. Finally we mention that, in general, even an elastically
symmetric resonance can be rectifying if screening is asymmetric.

\subsubsection{AC-Response}
\label{AC-Response}
\begin{figure}[t]
\hspace{1mm}
 \epsfxsize = 57 mm
 \epsffile{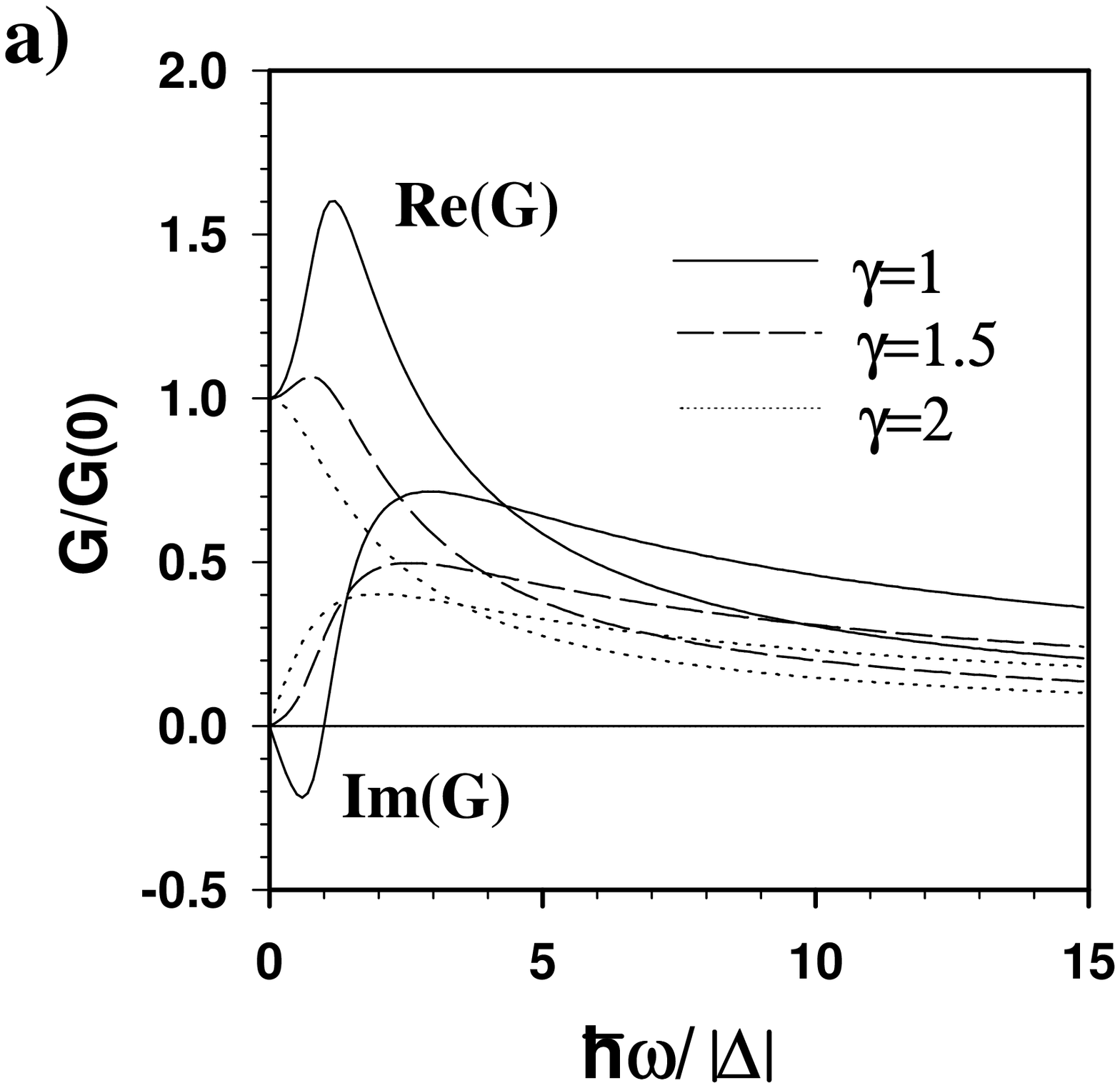}
\hspace{5mm}
 \epsfxsize = 57 mm
 \epsffile{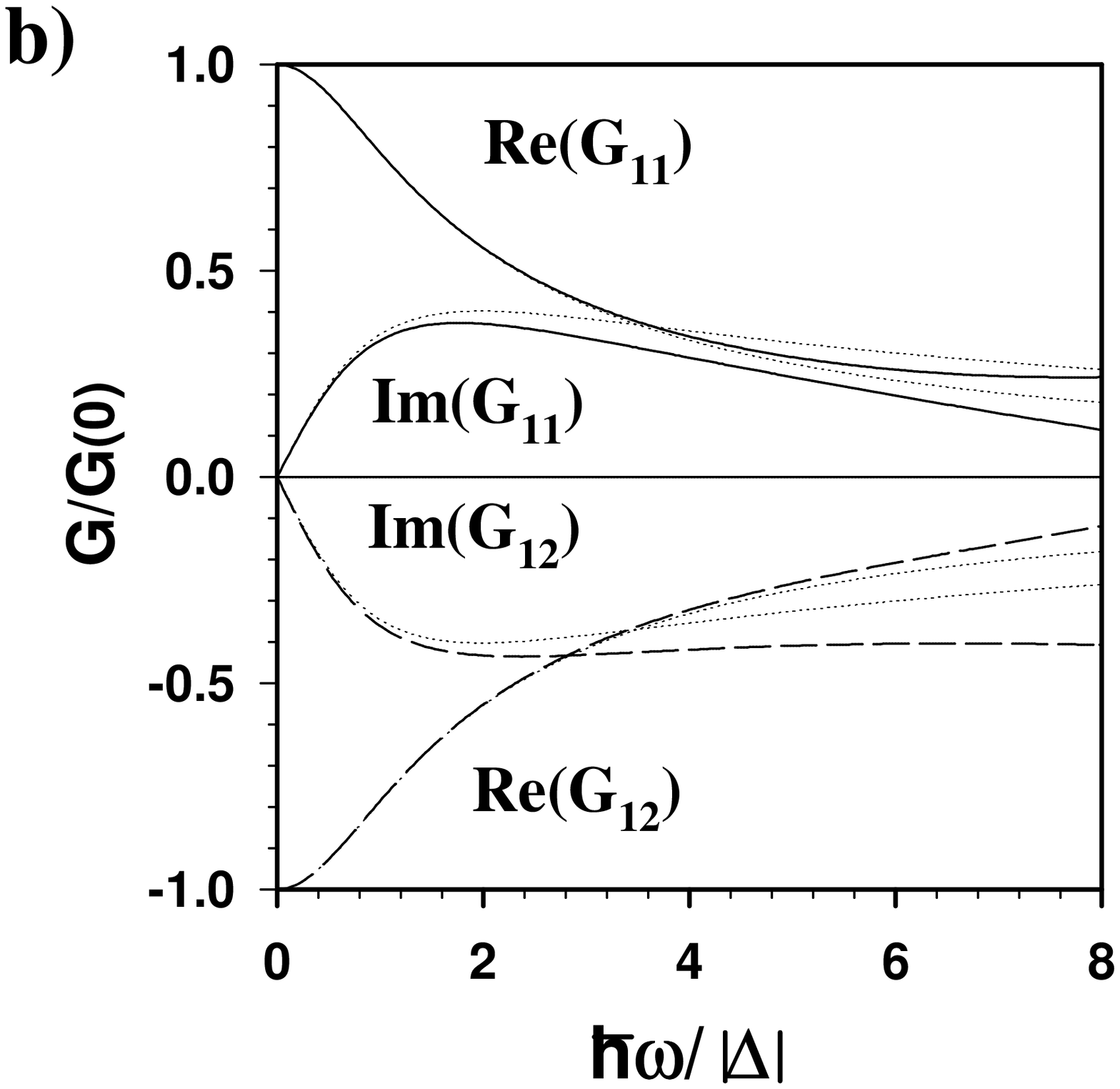}
 \vspace{-0.5cm}
\caption{ Frequency dependent admittance of the symmetric RTB.
a) charge neutral ($C=0$) case, $\gamma =  \Gamma /|\Delta |$;
b) Influence of a weak gate: $G_{11}$ (solid) and $G_{12}$ (dashed) for
$C|\Delta |/e^{2}=0.01$, $T=1$
(dotted: $C=0$).}
\label{rtbfig2}
\end{figure}

In order to calculate the admittances (\ref{eq61}) we must know the external response
$G_{\alpha \beta}^{e}$ defined for fixed potential. Using the specific scattering matrix elements
(\ref{rtb1}), Eq. (\ref{eq56}) gives
\begin{eqnarray}
G^{e}_{\alpha \beta} (\omega) = &
\frac{e^{2}}{h}\int dE\: \frac{f(E)-f(E+\hbar \omega)}{\hbar \omega}
\left( \delta _{\alpha \beta} -s_{\alpha \beta}^{\dagger }(E)s_{\alpha \beta}(E+\hbar \omega)\right)  \nonumber \\
 &  = a_{\alpha \beta} \; g_{I}(\omega ) 
\label{gextkurz}
\end{eqnarray} 
with
\begin{equation}
a_{\alpha \beta }(\omega) = \frac{4}{\Gamma (\hbar \omega + i\Gamma)}
\left( \begin{array}{cc}
\Gamma _{1} (\hbar \omega + i\Gamma _{2})  & -i \Gamma _{1} \Gamma _{2}   \\
-i \Gamma _{1} \Gamma _{2}   &\Gamma _{2} (\hbar \omega + i\Gamma _{1})
\end{array} \right)   
\label{aij}
\end{equation}
and where
\begin{equation}
g^{I}(\omega) = \frac{e^{2}}{h} \frac{i\Gamma}{4} \int dE\: \frac{f(E)-f(E+\hbar \omega)}{\hbar \omega}
\left( \frac{1}{\Delta (E+\hbar \omega) }-\frac{1}{\Delta ^{\ast}(E) }\right) \;\; .
\label{gi0}
\end{equation}
Here, we used
\begin{equation}
\frac{1}{\Delta (E+\hbar \omega)\Delta ^{\ast}(E)} =
\frac{1}{\hbar \omega + i\Gamma}\left(
\frac{1}{\Delta (E+\hbar \omega)} -\frac{1}{\Delta ^{\ast}(E)} \right) \;\;.
\label{zwire}
\end{equation}
It turns out that $g^{I}(\omega )$ is the ac-admittance for the symmetric barrier without gate ($C=0$). This is
is readily verified since for $C=0$ and for $a_{11}=a_{22}$ it holds
\begin{equation}
G (\omega)\equiv G^{I}_{11}(\omega )=
\frac{ G^{e}_{1 1} G^{e}_{22} - (G^{e}_{1 2})^{2}}
{ G^{e}_{1 1} + 2G^{e}_{1 2} + G^{e}_{22}}
=T_{max}\: g^{I}(\omega ) \;\; .
\label{gi}
\end{equation}
In the case of $C=0$ it is thus sufficient to discuss the symmetric barrier.
We consider first this case and assume zero temperature. The integral over the
difference of the Fermi functions reduces then to a simple integral from $\mu ^{(eq)}-\hbar \omega $
to $\mu ^{(eq)}$ and can be carried out analytically. We find
for the real part and the imaginary part of the admittance
\begin{eqnarray}
{\rm Re} [ g^{I} (\omega)] & = &
\frac{e^{2}}{h}\: \frac{\Gamma}{4\hbar \omega} \left( \arctan (\frac{\Delta E + \hbar \omega}{(\Gamma /2)})
- \arctan (\frac{\Delta E - \hbar \omega}{(\Gamma /2)}) \right)
\label{reg} \\
{\rm Im} [ g^{I} (\omega)] & = &
\frac{e^{2}}{h}\: \frac{\Gamma}{4\hbar \omega} \ln \sqrt{
\frac{(|\Delta |^{2}-(\hbar \omega)^{2})^{2}+(\Gamma \hbar \omega)^{2}}{|\Delta |^{2}}  }
\label{img}
\end{eqnarray}
where $|\Delta |\equiv |\Delta (\mu ^{(eq)})| $. Our results for $g^{I}(\omega)$ are equivalent
to earlier results of Fu and Dudley \cite{FUYD93a}. However, these authors neglect
Coulomb interaction and, coincidentally, for a symmetric RTB,
obtain the right expression only since they apply an antisymmetric bias $\pm \Delta V/2$ at the
contacts.\\ \indent
An expansion of ${\rm Re} [ G (\omega)] $ and ${\rm Im} [ G (\omega)] $ with respect to
frequency yields the coefficients defined in Eq. (\ref{expansion})
\begin{eqnarray}
G^{(0)} & = & \frac{e^{2}}{h} T(E_{F})  \;\;, \\
E & = & T_{max}\: \frac{D_{0}}{4}
\left( \frac{|\Delta |^{2}- \Gamma ^{2}/2}{|\Delta |^{2}}  \right)  \;\;,\\
K & = & \: \frac{h e^{2}T(E_{F})}{(2\pi |\Delta |)^{2}}
\left( \frac{|\Delta |^{2}- \Gamma ^{2}/3}{|\Delta |^{2}}  \right)  \;\;.
\end{eqnarray}
We see, that $E$ and $K$ change sign. In particular, for the symmetric case
this occurs if the value of the transmission probability $T$ equals $1/2$ and $3/4$, respectively.
In Fig. \ref{rtbfig2}a we plotted the real part and the imaginary part of the
normalized admittance $g^{I}(\omega )/g^{I}(0)$ as a function of $ \hbar \omega /|\Delta|$
for three different cases $T=1$ (solid), $T=0.5625$ (dashed), and
$T=0.25$ (dotted). For $T<3/4$ there is a peak in the conductance which belongs
to the excitation of the resonance by an energy quantum $\hbar \omega \approx | \Delta |$.
We mention that the normalized admittance is determined by a
single parameter $\gamma = \Gamma / |\Delta |$.\\ \indent%
%
%
%

%
We discuss now the effect of the capacitance $C$.
In principle, by using the Eqs. (\ref{gextkurz}) it is possible to derive expressions for
the conductances (\ref{eq61}). Here, we do not display the lengthy formulas
but we rather present the results in a figure. A comparison of the admittance for $C=0$ (thin dotted curves)
and for finite $C$ (thick curves) is shown in Fig. \ref{rtbfig2}b for the resonant case $T=1$. One sees that for
the small-capacitance case shown in the figure
the Coulomb-coupling to the external gate enhances the capacitive part of
the admittance. Indeed, an expansion with respect to the capacitance yields
\begin{equation}
G_{\alpha \beta}^{I}(\omega) = (-1)^{\alpha +\beta} G(\omega)
-i\omega C \frac{\Gamma _{\alpha}\Gamma_{\beta}}{\Gamma ^{2}} +{\cal O }(C ^{2})\;\; .
\label{}
\end{equation}    
In particular, one concludes $\sum _{\alpha \beta}G_{\alpha \beta}= -i\omega C +{\cal O }(C ^{2})$
which follows also directly from the fact that
the admittance $\sum _{\alpha \beta}G_{\alpha \beta}^{I}$ can be understood as
a resistor (with an external admittance $\sum _{\alpha \beta}G_{\alpha \beta}^{e}$)
in series with a capacitor (with a capacitance $C$).\\

The characteristic time-scales of tunneling problems \cite{MBRL82}
are a subject of considerable interest. 
With the analysis given above
we are now able to identify characteristic times for the electric
problem. We remark that the electrical problem investigated here
leads to answers not for single electron motion, but for 
the collective charge dynamics. 
Consider the case of a RTB in the zero capacitance limit. 
From the expansion of the conductance up to quadratic order in frequency
we can identify two characteristic frequencies or time-scales. 
There exists a frequency $\omega_{1}$ at which the {\it magnitude}
of the dc-current and the displacement current are equal.
The frequency $\omega_{1}$ is thus determined by 
$\tau_{1} = |E|/G^{0}$, where $E$ is the emittance and $G^{0}$ the
dc-conductance. The time-scale $\tau_{1}$ is thus a generalization
of the $RC$-time. For the RTB we find that this time is 
equal to $\tau_{1} = \hbar/\Gamma$ at resoance and is also equal to 
$\tau_{1} = \hbar/\Gamma$ far away from resonance. 
It vanishes for a Fermi energy at which the transmission 
of a symmetric barrier
is equal to $1/2$. A second characteristic frequency is obtained 
by comparing the displacement current determined by the emittance $E$
with the second order dissipative term $K$. The second 
characteristic time $\tau_{2}$ is thus determined by 
$\tau_{2} = |K|/|E|$. At resonance this time is given by 
$\tau_{2} = (1/3) (\hbar/\Gamma)$. 
This time tends to zero for Fermi energies far away from resonance. 
It is singular at the Fermi energies at which the emittance vanishes
and it is zero at the Fermi energies for which $K$ is zero. 
We conclude by mentioning that for the case in which the capacitance is 
not zero, the above consideration has to be applied to each element of 
the admittance matrix separately. In analogy to the tunneling time
problem, where there exist characteristic times for traversal and reflection,
from left and from right, the electrical problem will then be 
characterized by a set of characteristic frequencies or time scales 
for each admittance element, related only by sum rules due to 
current conservation and gauge invariance.

\section{Conclusion}
In this work we presented a theory for the admittance and the
nonlinear transport of open (i.e. connected to different reservoirs) mesoscopic conductors.
We have emphasized the application of this theory to a number of systems of current interest,
like quantum wires, quantum point contacts, and resonant tunneling barriers.
Our emphasis has been to derive results, which even so they might
not be realistic in detail, nevertheless capture the essential physics.
Our results should be useful both for comparison with additional theoretical work 
and with experimental work.

Due to the limitations of space, we have not reviewed a number of closely related 
subjects. The approach discussed here has been applied to ac-transport in two-dimensional 
electron gases in high magnetic fields \cite{CHRI96a,BUTT96b} which is 
particularly 
interesting in the regime where transport is dominated by carrier motion along edge states 
and by charging and de-charging of edge states.
Of fundamental interest are Aharonov-Bohm effects in capacitance coefficients if one 
of the capacitor plates has the form of a ring \cite{BUTT94b}
or for rings between capacitor plates \cite{BUTT95a}.
As in dc-transport these interference effects are closely related to the sample  
specific fluctuations away from an average behavior. For a chaotic cavity 
capacitance fluctuations away from an ensemble averaged capacitance have been investigated 
by Gopar and Mello and one of the authors \cite{GOPA96a}.
The ac-response and its fluctuations 
of a chaotic cavity connected to two leads and coupled capacitively to a back 
gate has been investigated by Brouwer and one of the authors.
This system is particularly interesting since the ensemble averaged quantities 
exhibit weak localization corrections \cite{BROUW}. 
Since the weak-localization effect is not associated with a net charge accumulation
the ac-response exhibits in addition to a Coulomb charge relaxation pole also a 
pole for uncharged excitations. Sch\"{o}ller \cite{SCHO96a} has investigated 
the dynamic capacitance of a one-dimensional wire. The dynamic capacitance 
exhibits interesting structure due to the plasma-modes of the one-dimensional
wire. 
This list illustrates that there are many avenues to extend the work presented here.
The application of the theory
to more realistic models is a quite challenging undertaking but very likely
also full of rewards.

\subsection*{Acknowledgments}
This work was supported by the Swiss National Science Foundation under grant
Nr. 43966.\\

\subsection*{Reference}
This work is to be published in
"Mesoscopic Electron Transport", 
edited by L. Kowenhoven, G. Schoen and L. Sohn,
NATO ASI Series E, (Kluwer Ac. Publ., Dordrecht).
\\

%

\begin{thebibliography}{99}
%
\bibitem{BUTT93d} B\"{u}ttiker, M., (1993) J. Phys.: Condens. Matter {\bf 5}, 9631.
%
\bibitem{CHRI96c} Christen, T., and B\"{u}ttiker, M., (1996) Europhys. Lett. {\bf 35}, 523.
%
\bibitem{JACK}    Jackson, J. D, (1996) Am. J. of Physics {\bf 64}, 855.
%
\bibitem{BUTT93b} B\"{u}ttiker, M., Thomas, H., and Pr\^etre, A., (1993) 
                  Phys. Rev. Lett. {\bf 70}, 4114; Pr\^etre, A., Thomas, H., 
                  and B\"uttiker, M., (1996) Phys. Rev. B, Oct.
%
%
\bibitem{PAST92a} Pastawski, H., (1992) Phys. Rev. B{\bf 46}, 4053.
%
\bibitem{FUYD93a} Fu, Y. and Dudley, S. C.,  (1993) Phys. Rev. Lett. {\bf 71}, 466.
%
\bibitem{BUTT94a} B\"{u}ttiker, M., Thomas, H., and Pr\^etre, A., (1994) Z. Phys. B {\bf 94}, 133. 
%
\bibitem{BUTT95b} B\"{u}ttiker, M., (1995) in "Quantum Dynamics and Submicron Structures",                 
                  edited by H. Cerdeira, G. Sch\"on, and B. Kramer, (Kluwer Academic Publishers,
                  Dordrecht) p. 657 -672.
%
\bibitem{CHRI96a} Christen, T., and B\"{u}ttiker, M., (1996) Phys. Rev. B {\bf 53}, 2064. 
%
\bibitem{CHRI96b} Christen, T., and B\"{u}ttiker, M., (1996) Phys. Rev. Lett. {\bf 76}, 143.
%
\bibitem{IMRY86a} Imry, Y., (1986) in {\it Directions in Condensed Matter Physics},
                  edited by G. Grinstein  and G. Mazenko, (World Scientific Singapore) p. 101. 
%
\bibitem{BEEN91a} Beenakker, C. W. J., and van Houten, H., (1991) {\it Quantum transport
                  in semiconductor nanostructures}, edited by.
                  H. Ehrenreich and D. Turnbull (New York Academic Press).
%
\bibitem{BUOT}    Buot, F. A., Phys. Rep. (1993) {\bf 234}, 73.
%
\bibitem{DATT93a} Datta, S., (1993) {\it  Electronic Transport in Mesoscopic Conductors},
                  Cambridge University Press, 1995; Buot, F. A., Phys. Rep. {\bf 234}, 73.
%
\bibitem{LAND70a} Landauer, R., (1970) Philos. Mag. {\bf 21}, 863.
%
\bibitem{BUTT86a} B\"{u}ttiker, M., (1986a) Phys. Rev. Lett. {\bf 57}, 1761.
%
\bibitem{BUTT88b} B\"{u}ttiker, M., (1988b) IBM J. Res. Develop. {\bf 32}, 317.
%
\bibitem{LAND57a} Landauer, R., (1957) IBM J. Res. Develop. {\bf 1}, 223.
%
\bibitem{FREN30a} Frenkel, J.,  (1930) Phys. Rev. {\bf 36}, 1604. 
%
\bibitem{LEVI89a} Levinson, I. B.,  (1989) Sov. Phys. JETP {\bf 68}, 1257. 
%
\bibitem{CHEN94a} Chen,  W., Smith, T. P., B\"{u}ttiker, M., and Shayegan, M., (1994) 
                  Phys. Rev. Lett. {\bf 73}, 146. 
%
\bibitem{SOMM96a} Sommerfeld, P. K. H., van der Heijden, R. W., and Peeters, F. M.,  
                  (1996) Phys. Rev. B {\bf 53}, 13250.
%
\bibitem{FIEL96a} Field, M., et al.,  (1996) Phys. Rev. Lett. {\bf 77}, 350.
%
\bibitem{PIEP94a} Pieper, J. B. and Price, J. C., (1994) Phys. Rev. Lett. {\bf 72}, 3586. 
%
\bibitem{KOUW94a} Kouwenhoven, L. P., et al., (1994) Phys. Rev. Lett. {\bf 73}, 3443.
%
\bibitem{REZN95a} Reznikov, M., Heiblum, M., Shtrikman, H., and Mahalu, D., (1995) Phys. Rev. Lett. {\bf 75}, 3340.
%
\bibitem{HOFB}    Hofbeck, K.,  Genzer, J., Schomburg, E., Ignatov,  A. A., Renk, K. F.,
                  Pavel'ev, D. G., Koschurinov, Yu.,  Melzer, B., Ivanov, S. 
                  Schaposchnikov, S., Kop'ev, P. S., (1996)
                  Phys. Lett. A{\bf 218}, 349.
%
\bibitem{TABO94a} Taboryski, R., et al.,  (1994) Phys. Rev. B {\bf 49}, 7813.
%
\bibitem{GLAZ89a} Glazman, L. I., and Khaetskii, A. V.,  (1989) Europhys. Lett. {\bf 9}, 263.
%
\bibitem{PATE90a} Patel, N. K., et al.,  (1990) J. Phys.: Condens. Matter {\bf 2}, 7247.
%
\bibitem{KOUW89a} Kouwenhoven, L. P., et al.,  (1989) Phys. Rev. B {\bf 39}, 8040.
%
\bibitem{KLUK89a} Kluksdahl, N. C., et al., (1989) Phys. Rev. B {\bf 39}, 7720.
%
\bibitem{BUTT96a} B\"uttiker, M., and Christen, T., (1996) in
                  {\em Quantum Transport in Semiconductor Submicron Structures},
                  edited by B. Kramer, (Kluwer Academic Publishers,Dordrecht);
                  NATO ASI Series, Vol. {\bf 326}, 263 -291.
%
\bibitem{BUTT85a} B\"{u}ttiker, M., Imry, Y., Landauer R., and Pinhas, S.,  (1985)
                  Phys. Rev. B {\bf 31}, 6207.                 
%
%
\bibitem{BUTT92b} B\"{u}ttiker, M., (1992b) Phys. Rev. B {\bf 46}, 12485.
%
\bibitem{GASP96a} Gasparian, V. M., Christen, T., and  B\"{u}ttiker,  M., (1996) 
                  Phys. Rev. {\bf A} Oct.
%
\bibitem{BUTT93a} B\"{u}ttiker, M., Pr\^etre, A., and Thomas, H., (1993) 
                  Phys. Lett. A {\bf 180}, 364.
%
\bibitem{BUTT96b} B\"uttiker, M., and Christen, T., (1996) 
                 {\em Dynamic Conductance in Quantum Hall Systems}, to appear in 
                 `The application of high magnetic fields in semiconductor physics',
                 edited by G. Landwehr, (unpublished). cond-mat/9607051
%
\bibitem{BUTT92}  B\"{u}ttiker, M., (1992) Phys. Rev. B {\bf 45}, 3807.                      
%
\bibitem{VANW88a} van Wees, B. J., et al., (1988) Phys. Rev. Lett. {\bf 60}, 848.
%
\bibitem{WHAR88a} Wharam, D. A., et al.,  (1988) J. Phys. C: Solid State Phys. {\bf 21}, L209.
%
\bibitem{BUTT90c} M. B\"{u}ttiker, Phys. Rev. B {\bf 41}, 7906 (1990).                            
%
\bibitem{MILL53a} Miller, S. C., and Good, R. M., (1953) Phys. Rev. {\bf 91}, 174.
%
\bibitem{SHKL}    Chklovskii, D. B., Shklovskii, B. I., and Glazman, L. I.,  (1992) 
                  Phys. Rev. B {\bf 46}, 4026.
%
\bibitem{BRUD94a} Bruder, C., and Sch\"{o}ller, H., (1994) Phys. Rev. Lett. {\bf 72}, 1076.
%
\bibitem{SCHO95a} Hettler, M. H., and Schoeller, H., (1995) Phys. Rev. Lett. {\bf 74}, 4907.
%
\bibitem{STAF96a} Stafford, C. A., and Wingreen, N. S., (1996)
                  Phys. Rev. Lett. {\bf 76}, 1916.
%
\bibitem{STAF96b} Stafford, C. A., (1996) Phys. Rev. Lett. {\bf 77}, 2770. 
%
\bibitem{MBRL82}  Landauer, R., and Martin, Th., (1994)
                  Rev. Mod. Physics {\bf 66}, 217.
%
\bibitem{BUTT94b} B\"{u}ttiker, M., (1994) 
                  Physica Scripta, T {\bf 54}, 104.
%
\bibitem{BUTT95a} B\"{u}ttiker, M. and Stafford, C. A., (1996) 
                  Phys. Rev. Lett. {\bf 76}, 495.
%
\bibitem{GOPA96a} Gopar, V. A., Mello, P. A., and B\"{u}ttiker,  M., (1996) 
                  Phys. Rev. Lett. {\bf 77} 3005.
%
\bibitem{BROUW}   Brouwer, P. W., and  B\"{u}ttiker,  M., (unpublished). 
%
\bibitem{SCHO96a} Sch\"{o}ller, H., (1996) (unpublished correspondence).
%

%
\end{thebibliography}
\end{document}